\documentclass[aps,prb,superscriptaddress,showpacs,twocolumn]{revtex4}
\usepackage[hypertex]{hyperref}
\usepackage{amsmath,amsfonts,bm,graphicx,amssymb}
\newcommand{\W}{\mathcal{W}}

\newcommand{\G}{\mathcal{G}}

\newcommand{\PP}{\mathcal{P}}
\newcommand{\QQ}{\mathcal{Q}}
\newcommand{\RR}{\mathcal{R}}

\newcommand{\hatrho}{\hat{\rho}}
\newcommand{\hatgamma}{\hat{\gamma}}
\newcommand{\hatH}{\hat{H}}
\newcommand{\hatsigma}{\hat{\sigma}}
\newcommand{\llangle}{\langle\!\langle}
\newcommand{\rrangle}{\rangle\!\rangle}
\newcommand{\rlangle}{\rangle\!\langle}
\newcommand{\rrllangle}{\rrangle\!\llangle}

\include{NonMarkovFCS.bib}
\begin{document}

\title{Counting statistics of transport through Coulomb blockade nanostructures: \\
High-order cumulants and non-Markovian effects}
\author{Christian Flindt}
\affiliation{Department of Physics, Harvard University, 17 Oxford
Street, Cambridge, MA 02138, USA}
\author{Tom\'{a}\v{s} Novotn\'{y}}
\affiliation{Department of Condensed Matter Physics,
             Faculty of Mathematics and Physics, Charles University,
             Ke Karlovu 5, 12116 Prague, Czech Republic}
\author{Alessandro Braggio}
\affiliation{CNR-SPIN,
             Dipartimento di Fisica,
             Universit\`{a} di Genova,
             Via Dodecaneso 33, 16146 Genova, Italy}
\author{Antti-Pekka Jauho}
\affiliation{Dept.\ of Micro og Nanotechnology, DTU Nanotech, Technical University of Denmark, Building 345east, 2800 Kongens Lyngby, Denmark}
\affiliation{Aalto University, Department of Applied Physics,  P.\
O.\ Box 11100, FI-00076 AALTO, Finland}

\date{\today}

\begin{abstract}
Recent experimental progress has made it possible to detect in real-time single
electrons tunneling through Coulomb blockade nanostructures,
thereby allowing for precise measurements of the statistical
distribution of the number of transferred charges, the so-called full counting statistics.
These experimental advances call for a solid theoretical platform for
equally accurate calculations of distribution functions and their
cumulants. Here we develop a general framework for calculating
zero-frequency current cumulants of arbitrary orders for transport through nanostructures
with strong Coulomb interactions. Our recursive method can treat
systems with many states as well as non-Markovian dynamics. We
illustrate our approach with three examples of current
experimental relevance: bunching transport through a two-level
quantum dot, transport through a nano-electromechanical system
with dynamical Franck-Condon blockade, and transport through
coherently coupled quantum dots embedded in a dissipative
environment. We discuss properties of high-order cumulants as well as possible subtleties associated with non-Markovian dynamics.
\end{abstract}


% 02.50.Ey Stochastic processes
% 03.65.Yz Decoherence; open systems; quantum statistical methods
% 72.70.+m Noise processes and phenomena
% 73.23.Hk Coulomb blockade; single-electron tunneling

\maketitle

\section{Introduction}

Electron transport through nanoscale structures is a stochastic process due to the randomness of the individual tunneling events. Quantum correlations and electron-electron interactions can strongly influence the transport process and thus the statistics of transferred charges. Full counting statistics\cite{Levitov1993,Levitov1996,Nazarov2003} concerns the distribution of the number of transferred charge, or equivalently, all corresponding cumulants (irreducible moments) of the distribution. Conventional transport measurements have focused on the first cumulant, the mean current, and in some cases also the second cumulant, the noise.\cite{Blanter2000} Higher order cumulants, however, reveal additional information concerning a variety of physical phenomena, including quantum coherence, entanglement, disorder, and dissipation.\cite{Nazarov2003} For example, non-zero higher-order cumulants reflect non-Gaussian behavior. Counting statistics in mesoscopic physics has been a subject of intensive theoretical interest for almost two decades, but recently it has also gained considerable experimental interest: in a series of experiments,\cite{Reulet2003,Bomze2005,Bylander2005,Fujisawa2006,Gustavsson2006,Fricke2007,Timofeev2007,Gershon2007,Gabelli2009,Flindt2009,Gustavsson2009,Fricke2010,Fricke2010b} high order cumulants
and even the entire distribution function of transferred charge have been measured, clearly demonstrating that counting statistics now has become an important concept also in experimental physics.

The theory of counting statistics was first formulated by Levitov and Lesovik for non-interacting electrons using a scattering formalism.\cite{Levitov1993,Levitov1996} Subsequent works have focused on the inclusion of interaction effects in the theory.\cite{Nazarov1999,Kindermann2003} In one approach, Coulomb interactions are incorporated via Markovian (generalized) master equations as originally developed by Bagrets and Nazarov.\cite{Bagrets2003} This approach is often convenient when considering systems with strong interactions, e.g., Coulomb-blockade structures. More recent developments include theories for finite-frequency counting statistics,\cite{Emary2007} conditional counting statistics,\cite{Sukhorukov2007} connections to entanglement entropy\cite{Klich2009} and to fluctuation theorems,\cite{Foerster2008,Esposito2009} and extensions to systems with non-Markovian dynamics.\cite{Braggio2006,Flindt2008,Schaller2009} The last topic forms the central theme of this paper.

We have recently published a series of papers on counting statistics.\cite{Flindt2005,Braggio2006,Flindt2008} Previous methods for evaluating the counting statistics of systems described by master equations had in practice been limited to systems with only a few states, and in Ref.\ \onlinecite{Flindt2005} we thus developed techniques for calculating the first few cumulants of Markovian systems with many states, for example nano-electromechanical systems.\cite{Flindt2004} In Ref.\ \onlinecite{Braggio2006}, Braggio and co-workers generalized the approach by Bagrets and Nazarov by including non-Markovian effects that may arise for example when the coupling to the electronic leads is not weak. The methods presented in these papers were subsequently unified and extended in Ref.\ \onlinecite{Flindt2008}, where we presented a general approach to calculations of cumulants of arbitrary order for systems with many states as well as with non-Markovian dynamics. The aim of the present paper is to provide a detailed derivation and description of this method, which recently has been used in a number of works,\cite{Zedler2009,Emary2009a,Emary2009b,Urban2009,Lindebaum2009,Zhong2009,Dominguez2010} as well as to illustrate its use with three examples of current experimental relevance.

The paper is organized as follows: In Sec.\ \ref{sec:non-markov} we
introduce the generic non-Markovian generalized master equation
(GME) which is the starting point of this work. The GME describes the
evolution of the reduced density matrix of the system, which has
been resolved with respect to the number of transferred particles.
Memory effects due to the coupling to the
environment as well as initial system-environment correlations are included in the GME.
Within this framework it is possible to calculate the finite-frequency current noise for non-Markovian GMEs\cite{Flindt2008} as we will discuss in future works. Section \ref{sec:non-markov} concludes with details of the superoperator notation used throughout the paper.

In Sec.\ \ref{sec:zerofreqcumu} we develop a theory for the
zero-frequency cumulants of the current. The cumulant generating
function (CGF) is determined by a single dominating pole of the
resolvent of the memory kernel, and its derivatives with respect to
the counting field evaluated at zero counting field yield the
cumulants of the current. Even in the Markovian case it is difficult
to determine analytically the dominating pole and in many cases one
would have to find it numerically. Numerical differentiation, however, is notoriously unstable,
and often one can only obtain accurate results for the first few
derivatives with respect to the counting field, i.\ e.\ the cumulants. In
order to circumvent this problem, we develop a numerically stable
recursive scheme based on a perturbation expansion in the counting
field. The scheme enables calculations of zero-frequency current
cumulants of very high orders, also for non-Markovian systems. Some
notes on the evaluation of the cumulants are presented, with the
more technical numerical details deferred to App.\ \ref{app:QR}.

Section~\ref{sec:univosc} gives a discussion of the generic behavior of high-order cumulants. As some of us have recently shown,\cite{Flindt2009} the high-order cumulants for basically any system (with or without memory effects) are expected to grow factorially in magnitude with the cumulant order and oscillate as functions of essentially any parameter as well as of the cumulant order. We describe the theory behind this prediction which is subsequently illustrated with
examples in Sec.\ \ref{sec:Markexamples}.

Section~\ref{sec:Markexamples} is devoted to two Markovian
transport models of current research interest, which we use to
illustrate our recursive scheme and the generic behavior of high-order cumulants discussed in Sec.\ \ref{sec:univosc}. We start with a model of
transport through a two-level quantum dot developed by
Belzig.\cite{Belzig2005} Due to the relatively simple analytic
structure of the model, it is possible to write down a closed-form
expression for the CGF, allowing us to develop a thorough
understanding of the behavior of high-order cumulants obtained
using our recursive scheme. We study the large deviation function of the system,\cite{Touchette2009}
which describes the tails of the distribution of
measurable currents, and discuss how it is related to the cumulants.

The second example concerns charge transport coupled to quantized
mechanical vibrations as considered in a recent series of papers on
transport through single
molecules\cite{Boese2001,Braig2003,McCarthy2003,Millis2004,Koch2005,Koch2005b,Pistolesi2008}
and other nano-electromechanical
systems.\cite{Gorelik1997,Armour2002,Fedorets2002,Novotny2003,Novotny2004,Flindt2004,Fedorets2004,Haupt2006,Rodrigues2006,Huebener2007,Harvey2008,Koerting2009,Huebener2009,Cavaliere2009prep,Harvey2009}
Due to the many oscillator states participating in transport the matrix
representations of the involved operators are of large dimensions
and it is necessary to resort to numerics. We demonstrate the
numerical stability of our recursive algorithm up to very high
cumulant orders ($\sim 100$) and show how
oscillations of the cumulants can be used to extract information
about the analytic structure of the cumulant generating function.
We calculate the large deviation function and show that it is highly sensitive to the
damping of the vibrational mode.

Section~\ref{sec:nonMarkov} concerns the counting statistics of non-Markovian systems. We consider  a model of non-Markovian electron
transport through a Coulomb-blockade double quantum dot embedded in
a dissipative heat bath and coupled to electronic leads. The dynamics of the charge populations of the double dot
can be described using a non-Markovian GME whose detailed derivation
is presented in App.\ \ref{app:doubledot}. We study the behavior
of the first three cumulants thus extending previous studies that
have been restricted to the noise.\cite{Aguado2004a,Aguado2004b} We
focus in particular on the influence of
decoherence\cite{Kiesslich2007} on the charge transport statistics. Finally, we discuss possible subtleties associated with non-Markovian dynamics and we provide the
reader with a unifying point of view on a number of results reported
in previous studies as well as in the examples discussed in this
paper.

Our conclusions are stated in Sec.~\ref{sec:conclusions}.
Appendix \ref{app:QR} describes  the numerical algorithms
used to solve the recursive equations for high-order cumulants,
while Apps.\ \ref{app:vibmol} and \ref{app:doubledot} give detailed derivations of
the Markovian GME for the vibrating molecule and the non-Markovian GME for the double-dot system, respectively.

\section{Generalized Master equation}
\label{sec:non-markov}

\begin{figure}
\begin{center}
\includegraphics[width=0.45\textwidth]{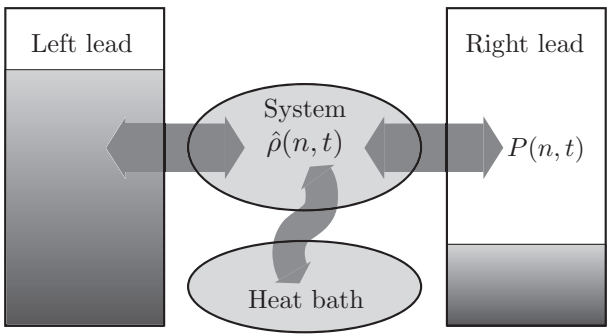}
\caption{Generic transport setup. A quantum system is connected to
electronic leads and a heat bath. A bias difference between the
leads drives electrons through the system, which can exchange energy
with the surrounding heat bath. The system is described by the
$n$-resolved density matrix $\hatrho(n,t)$ (see text), where $n$ is
the number of electrons that have been collected in the right lead
during the time span $[0,t]$. The probability distribution of $n$ is
denoted as $P(n,t)$.} \label{fig:gensetup}
\end{center}
\end{figure}

The generic transport setup under consideration in this work is
depicted in Fig.\ \ref{fig:gensetup}: A nanoscopic quantum system is
connected by tunneling barriers to two electronic leads, allowing
for charge and energy exchange with the leads. Typically, the
quantum system consists of a discrete set of (many-body) quantum
states. Moreover, the system is coupled to an external heat bath to
and from which energy can flow. We will be considering a transport
configuration, where a bias difference between the leads drives
electrons through the system.

The quantum system is completely described by its (reduced) density
matrix $\hatrho(t)$, obtained by tracing out the environmental degrees of freedoms, i.\ e.\ the electronic leads and the heat bath. It is, however, advantageous to resolve
$\hatrho(t)$ into the components $\hatrho(n,t)$, corresponding to
the number of electrons $n$ that have tunneled through the system
during the time span $[0,t]$.\cite{Makhlin2001,Shelankov2003,Wabnig2005} The $n$-resolved density matrix
allows us to study the statistics of the number of transferred
charges, similarly to well-known techniques from quantum optics.\cite{Cook1981,Lenstra1982,Plenio1998} We note that the (un-resolved) density matrix can always be recovered by summing over $n$, $\hatrho(t)=\sum_n\hatrho(n,t)$. For bi-directional processes, the number of tunneled electrons $n$ can be both positive and negative.

The major focus in the literature has been on systems obeying
Markovian dynamics;\cite{Blum1996,Gardiner2008,Alicki2007} however
recent years have witnessed an increased interest in non-Markovian
processes as well.\cite{Wilkie2000,Aissani2003,Breuer2006,Bellomo2007,Breuer2007,Budini2008,Timm2008,Breuer2009} In this
spirit we consider a generic non-Markovian generalized master
equation (GME) of the form
\begin{equation}
\frac{d}{dt}\hatrho(n,t)=\sum_{n'}\int_{0}^{t}dt'\W(n-n',t-t')\hatrho(n',t')+\hatgamma(n,t),
\label{eq:GME}
\end{equation}
obtained by tracing out the electronic leads and the heat bath. An
equation of this type arises for example in the partitioning scheme
devised by Nakajima and Zwanzig,\cite{Zwanzig2001} and in the
real-time diagrammatic technique for the dynamics of the reduced
density matrix on the Keldysh contour as described in Refs.\
\onlinecite{Schoeller1994,Konig1996,Konig1996b,Makhlin2001,Braggio2006}. The
memory kernel $\W$ accounts for the dynamics of the system
taking into account the influence of the degrees of freedom that
have been projected out, e.\ g.\ the electronic leads and the heat
bath. Here we have assumed that the
system is not explicitly driven by any time-varying fields, such
that the kernel $\W$ only depends on the time difference $t-t'$.
Additionally, we assume that the number of electrons $n$ that have
been collected in the right lead does not affect the system
dynamics, and the kernel consequently only depends on the difference
$n-n'$. The generic non-Markovian GME also contains the
inhomogeneity $\hatgamma(n,t)$ which accounts for initial
correlations between system and
environment.\cite{Zwanzig2001,Flindt2008} Typically, $\W(n,t)$ and
$\hatgamma(n,t)$ decay on comparable time scales, and $\hatgamma(n,t)$  thus vanishes in the long-time limit of Eq.\ (\ref{eq:GME}).

At this point, we note that while our method for extracting cumulants works
for {\it any} GME which satisfies certain, rather general conditions
specified in detail in Sec.~\ref{sec:zerofreqcumu},
the physical meaningfulness of the results nevertheless depends crucially on a
consistent derivation of the $n$-resolved memory kernel $\W(n,t)$. In Sec.\ \ref{subsec:nonMarkcorr} we discuss various subtleties associated with a proper derivation of the memory kernel for non-Markovian systems. In this work we use the notion of the Markovian limit of a general
non-Markovian GME in a somewhat loose manner, namely by referring to
the ``Markovian" limit of Eq.\ (\ref{eq:GME}) as the case, where
$\W(n,t)=\W(n)\delta(t)$ and $\hatgamma(n,t)=0$. We use this
terminology for the ease of notation, although we are aware that the
proper Markovian limit under certain circumstances may actually be
different. For an example of this, we refer the reader to
Ref.~\onlinecite{Spohn1979}, where it is demonstrated that the
correct Markovian limit for weak coupling theories should be
performed in the interaction picture. Since this procedure only
influences the off-diagonal elements in the weak coupling regime, we
ignore this subtlety in the rest of the paper as we will not be
considering such cases. In relevant situations this difference
should be taken into account --- it would, however, only lead to a
reinterpretation of the non-Markovian corrections studied in
Sec.~\ref{subsec:nonMarkcorr}. The issue of non-Markovian behavior,
its nature and distinction from Markovian approximations, is a
nontrivial and timely topic\cite{Wolf2008,Breuer2009,Rivas2009} which we only touch
upon briefly in this work, but our formalism paves the way for systematic studies of
such problems in the context of electronic noise and counting
statistics, for example as in Ref.~\onlinecite{Zedler2009}.

\subsection{Counting statistics}

In the following we introduce the
notion of cumulants of the charge transfer probability distribution,
and derive a formal expression for the cumulant generating function
CGF from the GME (\ref{eq:GME}). The probability
distribution for the number of transferred particles is obtained
from  the $n$-resolved density by tracing over the system degrees of
freedom,
\begin{equation}
P(n,t)=\mathrm{Tr}\{\hatrho(n,t)\}.
\end{equation}
Obviously, probability must be conserved, such that $\sum_n P(n,t)=1$. In order to study the cumulants of
$P(n,t)$ it is convenient to introduce a cumulant generating function
(CGF) $S(\chi,t)$ via the definition
\begin{equation}
e^{S(\chi,t)}\equiv\sum_{n}P(n,t)e^{in\chi},
\end{equation}
from which the cumulants $\llangle n^m\rrangle$ follow as
derivatives with respect to the counting field $\chi$ at $\chi=0$,
\begin{equation}
\llangle n^m\rrangle(t)\equiv\left.\frac{\partial^mS(\chi,t)}{\partial(i\chi)^m}\right|_{\chi\rightarrow0}.
\end{equation}
Alternatively, one can write
\begin{equation}
e^{S(\chi,t)}=\mathrm{Tr}\{\hatrho(\chi,t)\},
\end{equation}
which defines the $\chi$-dependent density matrix
\begin{equation}
\hatrho(\chi,t)\equiv\sum_{n}\hatrho(n,t)e^{in\chi}.
\end{equation}
By going to Laplace space via the transformation
\begin{equation}
\hatrho(\chi,z)\equiv\int_0^{\infty}dt\hatrho(\chi,t)e^{-zt},
\end{equation}
Equation (\ref{eq:GME}) transforms to an algebraic equation reading
\begin{equation}
z\hatrho(\chi,z)-\hatrho(\chi,t=0)=\W(\chi,z)\hatrho(\chi,z)+\hatgamma(\chi,z).
\label{eq:EOMinLaplace}
\end{equation}
This equation can be solved formally by introducing the resolvent
\begin{equation}
\mathcal{G}(\chi,z)\equiv[z-\W(\chi,z)]^{-1}.
\end{equation}
and writing
\begin{equation}
\hatrho(\chi,z)=\mathcal{G}(\chi,z)[\hatrho(\chi,t=0)+\hatgamma(\chi,z)].
\end{equation}
Finally, inverting the Laplace transform using the Bromwich integral
we obtain for the CGF\cite{Flindt2008}
\begin{equation}
e^{S(\chi,t)}=\frac{1}{2\pi
i}\int_{a-i\infty}^{a+i\infty}\!\!\!\!\!\!\!\!
dz\,\mathrm{Tr}\{\mathcal{G}(\chi,z)
[\hatrho(\chi,t=0)+\hatgamma(\chi,z)]\} e^{zt}, \label{eq:CGF2}
\end{equation}
where $a$ is larger than the real parts of all singularities of the
integrand.

Equation (\ref{eq:CGF2}) is a powerful formal result for the CGF, and, as we shall see,
it also leads to practical schemes for calculating current fluctuations.
In this work, we concentrate on the zero-frequency cumulants,
determined by the long-time limit of the CGF. The case of
finite-frequency noise,\cite{Flindt2008} where the inhomogeneity
$\hatgamma(\chi,z)$ plays an important role, will be considered
in future works.

\subsection{Notational details}

Throughout this paper we will use the superoperator notation
previously described in Ref.\ \onlinecite{Flindt2004}
and also used in a number of other
works.\cite{Flindt2005,Flindt2005b,Huebener2007,Flindt2008,Brandes2008,Harvey2008,Koerting2009,Huebener2009,Emary2009a,Emary2009b,Schaller2009,Harvey2009,Urban2009,Lindebaum2009,Zhong2009,Wu2010,Dominguez2010}
Using this notation, standard linear algebra operations
can conveniently be performed, analytically and numerically. Within
the formalism, the memory kernel $\W$, the resolvent $\G$, and other
operators that act linearly on density matrices, are referred to as
superoperators and denoted by calligraphic characters. Conventional
quantum mechanical operators, like the density matrix $\hatrho$,
acting in the conventional quantum mechanical Hilbert space, can be
considered themselves to span a Hilbert space, referred to as the
superspace. The superoperators act in the superspace, while
conventional quantum mechanical operators are considered as vectors
using a bra(c)ket notation, i.e.,
$\hat{V}\leftrightarrow|v\rrangle$, where $\hat{V}$ is a
conventional quantum mechanical operator, and $|v\rrangle$ is the
corresponding ket in the superspace. Double angle brackets are used
here in order to avoid confusion with conventional kets. In
numerical calculations, bras and kets are represented by vectors,
while superoperators are represented by matrices. The inner product
between bras and kets is defined as $\llangle
v|u\rrangle\equiv\mathrm{Tr}\{\hat{V}^{\dagger}\hat{U}\}$. Since the
involved superoperators, like $\W$ and $\G$, are not hermitian,
their eigenvalues are generally complex. In such cases, left and
right eigenvectors corresponding to a particular eigenvalue are not
related by hermitian conjugation. The left eigenvector, or bra,
corresponding to an eigenvalue $\lambda_k$ is therefore denoted with
a tilde, e.g.\ $\llangle \tilde{\lambda}_k|$, to avoid confusion
with the hermitian conjugate $|\lambda_k\rrangle^{\dagger}$ of the corresponding right eigenvector,
or ket, $|\lambda_k\rrangle$.

\section{Zero-frequency current cumulants}
\label{sec:zerofreqcumu}

In this section we derive the recursive method for evaluating
the zero-frequency current cumulants. We first define the zero-frequency
cumulants of the current as
\begin{equation} \llangle
I^m\rrangle\equiv\left.\frac{d}{dt}\llangle
n^m\rrangle(t)\right|_{t\rightarrow\infty}=\left.\frac{d}{dt}\frac{\partial^m
S(\chi,t)}{\partial(i\chi)^m}\right|_{\chi\rightarrow0,t\rightarrow\infty},
\label{eq:zerofreqcum}
\end{equation}
where $m=1,2,\ldots$.  As we shall show below, the
cumulants of the passed charge become linear in $t$ at long times
such that $\llangle n^m\rrangle(t)\rightarrow \llangle I^m\rrangle
t$, and the zero-frequency current cumulants are thus intensive
quantities (with respect to time). Thus, in the long-time limit
$\llangle I^m\rrangle/\llangle I\rrangle=\llangle
n^m\rrangle/\llangle n\rrangle$, and we  use these two normalized
quantities interchangeably throughout the paper.

In order to find the long-time limit of the CGF, we consider the
formal solution  Eq.\ (\ref{eq:CGF2}). The memory kernel
$\W(\chi,z)$ is assumed to have a single isolated eigenvalue
$\lambda_0(\chi,z)$, which for $\chi,z=0$ is zero, corresponding to
the stationary limit of $\hatrho(t)$, i.e.,
$\hatrho(t)\rightarrow\hatrho^{\mathrm{stat}}$ for large $t$. Here,
$\hatrho^{\mathrm{stat}}$ is the normalized solution to
$\W(\chi=0,z=0)\hatrho^{\mathrm{stat}}=0$. We exclude cases, where
the zero-eigenvalue is degenerate due to two or more uncoupled
sub-systems.\cite{vankampen2007} In the bracket notation
$\hatrho^{\mathrm{stat}}$ is denoted as $|0\rrangle$. The
corresponding left eigenvector can be found be noting that the
memory kernel with $\chi=0$ conserves probability for any $z$.
This can be inferred from the GME in Laplace space: For normalized
density matrices with $\mathrm{Tr}\{\hatrho(\chi=0,t)\}=1$, we have
$\mathrm{Tr}\{\hatrho(0,z)\}=1/z$, and Eq.\ (\ref{eq:EOMinLaplace})
yields
\begin{equation}
\mathrm{Tr}\{\W(0,z)\hatrho(0,z)\}+\mathrm{Tr}\{\hat{\gamma}(0,z)\}=0.
\label{eq:probconserv}
\end{equation}
It is generally possible to choose an initial state such that $\mathrm{Tr}\{\hat{\gamma}(0,z)\}=0$. The kernel does not depend on the choice of initial state and since Eq.\ (\ref{eq:probconserv}) holds for any normalized density matrix $\hatrho(0,z)$ we
deduce that $\mathrm{Tr}\{\W(0,z)\,\bullet\}=0$. In the bracket notation this equality can be expressed as $\llangle\tilde{0}|\W(0,z)=0$
with the left eigenvector $\llangle\tilde{0}|$ in the superspace
corresponding to the identity operator $\hat{1}$ in the conventional
Hilbert space. This moreover implies that\cite{Braggio2006}
\begin{equation}
\lambda_0(0,z)=0 \,\,\mathrm{for\,\, all}\,\, z.
\label{eq:lambda0}
\end{equation}

We next examine the
eigenvalue $\lambda_0(\chi,z)$ which we assume evolves adiabatically
from $\lambda_0(0,0)=0$ with small $\chi$ and $z$.  It is convenient
to introduce the  mutually orthogonal projectors
\begin{equation}
\PP(\chi,z)=\PP^2(\chi,z)=|0(\chi,z)\rrllangle\tilde{0}(\chi,z)|
\end{equation}
and
\begin{equation}
\QQ(\chi,z)=\QQ^2(\chi,z)\equiv 1-\PP(\chi,z)
\end{equation}
with $\PP(\chi,z)$ developing adiabatically from
$\PP(0,0)\equiv|0\rrllangle\tilde{0}|$ for small $\chi$ and $z$. Here, $\llangle\tilde{0}(\chi,z)|$ and $|0(\chi,z)\rrangle$ are the
left and right eigenvectors corresponding to $\lambda_0(\chi,z)$,
which develop adiabatically from $\llangle\tilde{0}|$ and
$|0\rrangle$, respectively.  In
terms of $\PP(\chi,z)$ and $\QQ(\chi,z)$ the memory kernel can be
partitioned as
\begin{equation}
\W(\chi,z)=\lambda_0(\chi,z)\PP(\chi,z)+\QQ(\chi,z)\W(\chi,z)\QQ(\chi,z).
\label{eq:parti1}
\end{equation}
In deriving this expression we used
\begin{equation}
\PP(\chi,z)\W(\chi,z)\PP(\chi,z)=\lambda_0(\chi,z)\PP(\chi,z).
\end{equation}
Using the partitioning, Eq.\ (\ref{eq:parti1}), the resolvent becomes
\begin{equation}
\G(\chi,z)=\frac{\PP(\chi,z)}{z-\lambda_0(\chi,z)}+\QQ(\chi,z)\frac{1}{z-\W(\chi,z)}\QQ(\chi,z).
\label{eq:partioning}
\end{equation}
For $\chi=0$ the first term of the resolvent has a simple pole at
$z=0$, which determines the long-time limit, i.e. it corresponds to
the stationary state $\hatrho^{\mathrm{stat}}$. We denote the pole at
$z=0$ by $z_0$. All singularities of the second term have negative real
parts and do not contribute in the long-time limit. Again, we assume
adiabatic evolution of the pole $z_0(\chi)$ from $z_0(0)=0$ with
small $\chi$, such that $z_0(\chi)$ is the particular pole that
solves\cite{Braggio2006,Flindt2008}
\begin{equation}
z_0-\lambda_0(\chi,z_0)=0 \label{eq:pole}.
\end{equation}
With small $\chi$, the other
singularities still have more negative real parts and the pole
$z_0(\chi)$ again determines the long-time behavior. From Eq.\
(\ref{eq:CGF2}) we then find for large $t$
\begin{equation}
e^{S(\chi,t)}\rightarrow D(\chi,z_0)e^{z_0(\chi)t}, \label{eq:CGF3}
\end{equation}
where
\begin{equation}
D(\chi,z_0)=\mathrm{Tr}\{\PP(\chi,z_0)[\hatrho(\chi,t=0)+\hat{\gamma}(\chi,z_0)]\}.
\end{equation}
From the definition of the zero-frequency current cumulants in Eq.\
(\ref{eq:zerofreqcum}) we then establish that
\begin{equation}
z_0(\chi)=\sum_{n=1}^{\infty}\frac{(i\chi)^{n}}{n
 !}\llangle I^{n} \rrangle.
 \label{eq:poleCGF}
\end{equation}
We note that the CGF in the long-time limit and thus the
zero-frequency cumulants do not depend on the initial state
$\hatrho(\chi,t=0)$ and the inhomogeneity $\hat{\gamma}(\chi,z)$. In contrast, both $\hatrho(\chi,t=0)$ and $\hat{\gamma}(\chi,z)$ must be appropriately incorporated in order to calculate the finite-frequency noise.\cite{Flindt2008} Equations (\ref{eq:pole}) and (\ref{eq:poleCGF}) form the main
theoretical result of this section, generalizing earlier results for
Markovian systems.\cite{Bagrets2003,Flindt2005} In the Markovian
limit, the memory kernel and the corresponding eigenvalue close to 0
have no $z$-dependence, and Eq.\ (\ref{eq:pole}) immediately yields
$z_0(\chi)=\lambda_0(\chi)$,\cite{Bagrets2003,Flindt2005} where $\lambda_0(\chi)$ is the
eigenvalue of the $z$-independent kernel, which goes to zero with
$\chi$ going to zero, i.e. $\lambda_0(0)=0$.

Although, we have formally derived an expression for the CGF, it may
in practice, given a specific memory kernel $\W(\chi,z)$, be
difficult to determine the eigenvalue $\lambda_0(\chi,z)$ including
its dependence on $\chi$ and $z$. Moreover, the solution of Eq.\
(\ref{eq:pole}) itself poses an additional problem, which needs to
be addressed in the non-Markovian case. In the Markovian limit, only
derivatives of the eigenvalue $\lambda_0(\chi)$ with respect to the
counting field $\chi$ need to be determined. However, with the superoperator $\W(\chi)$ being represented by a matrix of size $N\times N$, there is no closed-form expression for the eigenvalue $\lambda_0(\chi)$ already with $N>4$. The immediate alternative
strategy would then be to calculate numerically the eigenvalue and the derivatives with respect to $\chi$.
Typically, however, this is a numerically unstable
procedure, which is limited to the first few
derivatives.\cite{Press2007} Consequently, we devote the rest of
this section to the development of a numerically stable, recursive
scheme that solves Eqs.\ (\ref{eq:pole}) and (\ref{eq:poleCGF}) for
high orders of cumulants, including in the non-Markovian case.

\subsection{Recursive scheme}

\subsubsection{The Markovian case}

We consider first the Markovian case,\cite{Flindt2005,Baiesi2009}
before proceeding with the general non-Markovian case. In the
Markovian case, the memory kernel $\W$ has no $z$-dependence, and
the current cumulants are determined by the eigenvalue
$\lambda_0(\chi)$ which solves the eigenvalue problem
\begin{equation}
\W(\chi)|0(\chi)\rrangle=\lambda_0(\chi)|0(\chi)\rrangle,
\label{eq:eigprob1}
\end{equation}
where $\lambda_0(0)=0$. We find the eigenvalue using perturbation
theory in the counting field $\chi$ in a spirit similar to that of
standard Rayleigh-Schr\"{o}dinger perturbation theory. To this end
we introduce the unperturbed operator $\W\equiv\W(0)$ and the
perturbation $\Delta\W(\chi)$ such that
\begin{equation}
\W(\chi)=\W+\Delta\W(\chi).
\end{equation}
We can then write
\begin{equation}
\lambda_0(\chi)=\llangle\tilde{0}|\Delta\W(\chi)|0(\chi)\rrangle,
\label{eq:eigenvalue}
\end{equation}
where we have used $\llangle\tilde{0}|\W=0$ and chosen the
conventional normalization $\llangle\tilde{0}|0(\chi)\rrangle=1$. We
moreover employ the shorthand notation
$\PP=\PP^2\equiv\PP(0,0)=|0\rrllangle\tilde{0}|$ and
$\QQ=\QQ^2\equiv 1-\PP$ for the projectors introduced in the
previous section, and write
\begin{equation}
|0(\chi)\rrangle=|0\rrangle+\QQ|0(\chi)\rrangle, \label{eq:vector}
\end{equation}
consistently with the choice of normalization. Using that
$\W=\QQ\W\QQ$, Eq.\ (\ref{eq:eigprob1}) can be written
\begin{equation}
\QQ\W\QQ|0(\chi)\rrangle=[\lambda_0(\chi)-\Delta\W(\chi)]|0(\chi)\rrangle.
\label{eq:QWQ}
\end{equation}
Next, we introduce the pseudo-inverse\cite{Flindt2004,Flindt2005}
defined as\footnote{We note that our definition of the
pseudo-inverse differs from the Moore-Penrose pseudo-inverse $\RR_{M\!P}$, which
is implemented in many numerical software packages, e.g., in Matlab.
However, when projected on the regular subspace by
$\QQ$ it reduces to our pseudo-inverse, i.e. $\QQ\RR_{M\!P}\QQ=\RR$.}
\begin{equation}
\RR=\QQ\W^{-1}\QQ.
\end{equation}
The pseudo-inverse is a well-defined object, since the inversion is
performed in the subspace corresponding to $\QQ$, where $\W$ is
regular. By applying $\RR$ on both sides of  Eq.\ (\ref{eq:QWQ}) we
find
\begin{equation}
\QQ|0(\chi)\rrangle=\RR[\lambda_0(\chi)-\Delta\W(\chi)]|0(\chi)\rrangle,
\end{equation}
which combined with Eq.\ (\ref{eq:vector}) yields
\begin{equation}
|0(\chi)\rrangle=|0\rrangle+\RR[\lambda_0(\chi)-\Delta\W(\chi)]|0(\chi)\rrangle.
\label{eq:eigenvector}
\end{equation}
 Equations (\ref{eq:eigenvalue}) and (\ref{eq:eigenvector}) form the basis of
the recursive scheme developed below.

We first Taylor expand the eigenvalue $\lambda_0(\chi)$, the
eigenvector $|0(\chi)\rrangle$, and the perturbation
$\Delta\W(\chi)$, around $\chi=0$ as
\begin{equation}
\begin{split}
\lambda_0(\chi)=&\sum_{n=1}^{\infty}\frac{(i\chi)^n}{n!}\llangle I^n\rrangle,\\
|0(\chi)\rrangle=&\sum_{n=0}^{\infty}\frac{(i\chi)^n}{n!}|0^{(n)}\rrangle,\\
\Delta\W(\chi)=&\sum_{n=1}^{\infty}\frac{(i\chi)^n}{n!}\W^{(n)},
\end{split}
\end{equation}
where we have used that $\lambda_0(0)=0$ and $\Delta\W(0)=0$. Inserting these expansions into
Eqs.\ (\ref{eq:eigenvalue}) and (\ref{eq:eigenvector}), and
collecting terms to same order in $\chi$, we arrive at a recursive
scheme reading
\begin{equation}
\begin{split}
\llangle I^n\rrangle_=&\sum_{m=1}^n{n\choose m}\llangle\tilde{0}|\W^{(m)}|0^{(n-m)}\rrangle,\\
|0^{(n)}\rrangle = &\mathcal{R}\sum_{m=1}^n{n \choose m}
\left[\llangle I^{m}\rrangle-\W^{(m)}\right]|0^{(n-m)}\rrangle,
\end{split}
\label{eq:recursivescheme}
\end{equation}
for $n=1,2,\ldots$. The recursive scheme allows for systematic calculations of
cumulants of high orders.

As illustrative examples we evaluate the first three current
cumulants using the recursive scheme,
\begin{equation}
\begin{split}
\llangle
I^1\rrangle_M=&\llangle\tilde{0}|\W^{(1)}|0\rrangle,\\
\llangle I^2\rrangle_M=&\llangle\tilde{0}|\left(\W^{(2)}-2\W^{(1)}\RR\W^{(1)}\right)|0\rrangle,\\
\llangle I^3\rrangle_M=&\llangle\tilde{0}|\left(\W^{(3)}+6\W^{(1)}\RR\W^{(1)}\RR\W^{(1)}\right.\\
&-3\{\W^{(2)}\RR\W^{(1)}+\W^{(1)}\RR\W^{(2)}\}\\
&\left.-6\llangle I^1\rrangle_M\W^{(1)}\RR^2\W^{(1)}\right) |0\rrangle,\\
\end{split}\label{eq:firstthree}
\end{equation}
having used
$|0\rrangle\equiv|0^{(0)}\rrangle$ and $\RR|0\rrangle=0$, since
$\QQ|0\rrangle=0$. The subscript $M$ reminds us that these results hold
for the Markovian case. The expressions (\ref{eq:firstthree}) for
the first three cumulants are equivalent to the ones derived in
Ref.\ \onlinecite{Flindt2005}, albeit using a slightly different
notation. Importantly, the recursive scheme presented here allows
for an easy generation of higher order cumulants, either
analytically or numerically.

\subsubsection{The non-Markovian case}

We now proceed with the non-Markovian case, where we first need to
consider the eigenvalue problem
\begin{equation}
\W(\chi,z)|0(\chi,z)\rrangle=\lambda_0(\chi,z)|0(\chi,z)\rrangle
\label{eq:eigprob2}
\end{equation}
where $\lambda_0(\chi,z)$ is the particular eigenvalue for which
$\lambda_0(0,z)=0$. The basic equations, Eqs.\ (\ref{eq:eigenvalue})
and (\ref{eq:eigenvector}), are still valid, provided that
$\lambda_0(\chi)$, $|0(\chi)\rrangle$, $\W$, and $\Delta\W(\chi)$
are replaced by $\lambda_0(\chi,z)$, $|0(\chi,z)\rrangle$,
$\W=\W(0,0)$, and $\Delta\W(\chi,z)=\W(\chi,z)-\W(0,0)$,
respectively, i.e.,
\begin{equation}
\lambda_0(\chi,z)=\llangle\tilde{0}|\Delta\W(\chi,z)|0(\chi,z)\rrangle,
\label{eq:eigenvalue2}
\end{equation}
and
\begin{equation}
|0(\chi,z)\rrangle=|0\rrangle+\RR[\lambda_0(\chi,z)-\Delta\W(\chi,z)]|0(\chi,z)\rrangle.
\label{eq:eigenvector2}
\end{equation}
Again, we Taylor expand all objects around $\chi=0$, but in this
case also around $z=0$,
\begin{equation}
\begin{split}
\lambda_0(\chi,z)=&\sum_{n,l=0}^{\infty}\frac{(i\chi)^n}{n!}\frac{z^l}{l!}c^{(n,l)},\\
|0(\chi,z)\rrangle=&\sum_{n,l=0}^{\infty}\frac{(i\chi)^n}{n!}\frac{z^l}{l!}|0^{(n,l)}\rrangle,\\
\Delta\W(\chi,z)=&\sum_{n,l=0}^{\infty}\frac{(i\chi)^n}{n!}\frac{z^l}{l!}\W^{(n,l)},
\end{split}
\label{eq:expansions}
\end{equation}
with $\W^{(0,0)}=0$ by definition and $c^{(0,l)}=0$, since
$\lambda_0(0,z)=0$. Inserting these expansions into Eqs.\
(\ref{eq:eigenvalue2}) and (\ref{eq:eigenvector2}) and collecting
terms to same orders in $\chi$ and $z$, we find a recursive scheme
reading
\begin{equation}
\begin{split}
c^{(n,l)}=&\sum_{m=1}^n{n\choose m}\sum_{k=0}^l{l\choose k}\llangle\tilde{0}|\W^{(m,k)}|0^{(n-m,l-k)}\rrangle,\\
|0^{(n,l)}\rrangle = &\mathcal{R}\sum_{m=0}^n\!{n \choose
m}\!\sum_{k=0}^l\!{l \choose
k}\!\left[c^{(m,k)}\!-\!\W^{(m,k)}\right]\!|0^{(n-m,l-k)}\rrangle.
\end{split}
\label{eq:recursivescheme1}
\end{equation}
In case the memory kernel has no $z$-dependence, corresponding to
the Markovian case, only terms with $l=0$ are non-zero, and the
recursive scheme reduces to the one given in Eq.\
(\ref{eq:recursivescheme}). In particular, the coefficients
$c^{(n,0)}$ equal the current cumulants $\llangle I^n\rrangle_M$ in
the Markovian limit of the kernel, $z\rightarrow 0$.

In the non-Markovian case, we need to proceed with the solution of
Eq.\ (\ref{eq:pole}) for $z_0$ and extract the current cumulants
$\llangle I^{n}\rrangle$. Inserting the expression for $z_0$ in Eq.\
(\ref{eq:poleCGF}) into Eq.\ (\ref{eq:pole}) and using the expansion
of $\lambda_0(\chi,z)$ given in Eq.\ (\ref{eq:expansions}), we find
\begin{equation}
\sum_{n=1}^{\infty}\frac{(i\chi)^{n}}{n !}\llangle
I^{n}\rrangle=\sum_{k,l=0}^{\infty}\frac{(i\chi)^k}{k!}\frac{1}{l!}\left\{\sum_{n=1}^{\infty}\frac{(i\chi)^{n}
}{n !}\llangle I^{n}\rrangle\right\}^lc^{(k,l)}.
\end{equation}
Collecting terms to same order in $\chi$, we find
\begin{equation}
\llangle
I^n\rrangle=n!\sum_{k,l=0}^n\frac{1}{k!}\frac{1}{l!}P^{(n-k,l)}c^{(k,l)},
\label{eq:recursivescheme2}
\end{equation}
in terms of the auxiliary quantity
\begin{equation}
P^{(k,l)}\equiv\sum_{\substack{n_1,\ldots,n_l=1\\
n_1+\ldots+n_l=k}}^{k}\frac{\llangle
I^{n_1}\rrangle}{n_1!}\cdots\frac{\llangle
I^{n_l}\rrangle}{n_l!},\,\, l\geq 1,
\end{equation}
where only terms in the sums for which $n_1+\ldots+n_l=k$ should be
included. For $l=0$, we have $P^{(k,0)}\equiv\delta_{k,0}$. The
auxiliary quantity can also be evaluated recursively by noting that
\begin{equation}
P^{(k,l)}=\sum_{n=1}^{k}\frac{\llangle
I^{n}\rrangle}{n!}P^{(k-n,l-1)}, \label{eq:recursivescheme3}
\end{equation}
with the boundary conditions $P^{(k,0)}=\delta_{k,0}$,
$P^{(0,l)}=\delta_{0,l}$, and $P^{(k,-1)}\equiv 0$.

When combined, Eqs.\ (\ref{eq:recursivescheme1},
\ref{eq:recursivescheme2}, \ref{eq:recursivescheme3}) constitute a
recursive scheme which allows for numerical or analytic calculations
of cumulants of high orders in the general non-Markovian case. As
simple examples, we show the first three cumulants\cite{Flindt2008} obtained from
Eqs.\ (\ref{eq:recursivescheme2}), (\ref{eq:recursivescheme3}), in
terms of the coefficients $c^{(n,l)}$
\begin{equation}
\begin{split}
\llangle
I^1\rrangle =& c^{(1,0)},\\
\llangle I^2\rrangle =& c^{(2,0)}+2c^{(1,0)}c^{(1,1)},\\
\llangle I^3\rrangle =& c^{(3,0)}+3 c^{(2,0)}c^{(1,1)}\\
&+3c^{(1, 0)}\left[c^{(1, 0)} c^{(1, 2)}+2(c^{(1,1)})^2+ c^{(2,1)}\right].\\
\end{split}
\label{eq:nonMarkovcumulants}
\end{equation}
In general, the $n$'th current cumulant $\llangle I^n\rrangle$
contains the coefficients
\begin{equation}
c^{(k,l)}=\partial_{(i\chi)}^k\partial_{z}^l\lambda_0(\chi,z)|_{\chi,z\rightarrow
0}
\end{equation}
with $1\leq k+l\leq n$. However, coefficients of the form
$c^{(0,l)}$ are zero since $\lambda_0(0,z)\equiv 0$ as discussed
below Eq.~\eqref{eq:probconserv} and it thus suffices to consider
$l\leq n-1$. From Eq.~\eqref{eq:recursivescheme1} it follows that
$c^{(k,l)}$ depend only on $\W^{(m,n)}$ with $m\leq k$ and $n\leq l$
so that we can conclude that the $n$'th cumulant of the current
depends at maximum on the $(n-1)$'th time-moment of the memory
kernel $\int_0^{\infty}dt \,t^{n-1} \W(\chi,t)$. In particular, this
implies that the mean current is a purely Markovian quantity
depending only on the time-integrated memory kernel while the second
and higher order cumulants deviate from the results in the Markovian
case.\cite{Braggio2006}

The coefficients $c^{(n,l)}$ can be found from Eq.\
(\ref{eq:recursivescheme1}). Coefficients of the form $c^{(n,0)}$ only contain zeroth order terms
in $z$ and are, as already mentioned, equal to the current cumulants
$\llangle I^n\rrangle_M$ in the Markovian limit, i.e.,
\begin{equation}
c^{(n,0)}=\llangle I^n\rrangle_M,\,\, n=1,2,3,\ldots.
\end{equation}
For the other coefficients entering the expressions in  Eq.\ (\ref{eq:nonMarkovcumulants}) for the first three non-Markovian current cumulants, we find
\begin{equation}
\begin{split}
c^{(1,1)}=&\llangle\tilde{0}|\left(\W^{(1,1)}-\W^{(1,0)}\RR\W^{(0,1)}\right)|0\rrangle,\\
c^{(1,2)}=&\llangle\tilde{0}|\left(\W^{(1,2)}-2\W^{(1,1)}\RR\W^{(0,1)}-\W^{(1,0)}\RR\W^{(0,2)}\right.\\
&\left.+2\W^{(1,0)}\RR\W^{(0,1)}\RR\W^{(0,1)}\right)|0\rrangle,\\
c^{(2,1)}=&\llangle\tilde{0}|\left(\W^{(2,1)}+2\W^{(1,0)}\RR\W^{(0,1)}\RR\W^{(1,0)}\right.\\
&+2\W^{(1,0)}\RR\W^{(1,0)}\RR\W^{(0,1)}-2\W^{(1,1)}\RR\W^{(1,0)}\\
&\left.-2\W^{(1,0)}\RR\W^{(1,1)}-\W^{(2,0)}\RR\W^{(0,1)}\right)|0\rrangle.\\
\end{split}
\label{eq:nonMarkovcoeff}
\end{equation}
Again, as in the Markovian case, higher order cumulants including the coefficients $c^{(k,l)}$ are readily
generated, analytically or numerically. The results presented here can be generalized to the
statistics of several different counted quantities as in Ref.\ \onlinecite{Sanchez2007,Sanchez2008b}, and  cross-correlations can be evaluated using the same compact notation developed in this work.\cite{Braggio2009b}

\subsection{Notes on evaluation}
\label{subsec:evaluation}

As previously mentioned, the size of the memory kernel $\W(\chi,z)$
could in practice hinder the calculation of $\lambda_0(\chi,z)$ and
the solution of Eq.\ (\ref{eq:pole}), and thus the evaluation of the
current cumulants. The recursive scheme described above,
however, only relies on the ability to solve matrix equations and
perform matrix multiplications. Both of these operations are
numerically feasible and stable, even when the involved matrices are of large
dimensions. In general, the recursive scheme requires the following
steps: The stationary state must be found by solving
\begin{equation}
\W|0\rrangle=0, \label{eq:hom_eq}
\end{equation}
with the normalization requirement
$\llangle\tilde{0}|0\rrangle=\mathrm{Tr}\{\hat{1}^{\dagger}\hatrho^{\mathrm{stat}}\}=1$.
Secondly, the $\chi$ and $z$ derivatives of the memory kernel must
be found
\begin{equation}
\W^{(n,l)}=\left.\partial^n_{(i\chi)}\partial^l_{z}\W(\chi,z)\right|_{\chi,z\rightarrow
0}
\end{equation}
for $(n,l)\neq (0,0)$. Typically, the dependence on the counting
field $\chi$ enters matrix elements in an exponential function (see
e.\ g.\ Refs.\ \onlinecite{Nazarov2003}, \onlinecite{Bagrets2003}, \onlinecite{Braggio2006} and
examples in Secs. \ref{sec:Markexamples} and \ref{sec:nonMarkov}), e.\ g.\
as a factor of $e^{i\chi}$, for which the derivatives with respect
to $\chi$ are easily found analytically. The $z$-dependence of the
matrix element $[\W(\chi,z)]_{kj}$ can be written
\begin{equation}
[\W(\chi,z)]_{kj}=\int_0^{\infty}dt [\W(\chi,t)]_{kj} e^{-zt},
\end{equation}
such that
\begin{equation}
[\W^{(n,l)}]_{kj}=\int_0^{\infty}dt (-t)^l
\left[\partial^n_{(i\chi)}\W(\chi,t)|_{\chi\rightarrow
0}\right]_{kj}.
\end{equation}
The integration over time can be performed in a numerically stable
manner for arbitrary $n$,\cite{Press2007} thereby avoiding taking
numerical derivatives with respect to $z$.

Finally, matrix multiplications have to be performed. Here, special
attention has to be paid to terms involving the pseudo-inverse
$\RR$, i.e.\ $\RR|x\rrangle$, where $|x\rrangle$ for example has the
form $\W^{(0,1)}|0\rrangle$ in the expression for the coefficient $c^{(1,1)}$ in Eq.\ (\ref{eq:nonMarkovcoeff}). In order to evaluate such expressions we introduce
$|y\rrangle$ as the solution(s) to
\begin{equation}
\W|y\rrangle=\QQ|x\rrangle, \label{eq:Req}
\end{equation}
such that
\begin{equation}
\QQ|y\rrangle=\RR|x\rrangle, \label{eq:RRx}
\end{equation}
which can be verified by applying $\RR$ on both sides of Eq.\
(\ref{eq:Req}) and using that $\RR\W=\QQ\W^{-1}\QQ\W=\QQ$ and
$\RR\QQ=\RR$. The projector $\QQ$ in Eq.\ (\ref{eq:Req}) ensures
that the right hand side lies in the range of $\W$, and since $\W$
is singular, the equation has infinitely many solutions. The
solutions can be written
\begin{equation}
|y\rrangle=|y_0\rrangle+c|0\rrangle,\,\, c\in\mathbb{C},
\end{equation}
where $|y_0\rrangle$ is a particular solution to Eq.\
(\ref{eq:Req}), which can be found numerically. We then obtain
$\RR|x\rrangle$ by applying $\QQ$ to $|y\rrangle$ according to Eq.\
(\ref{eq:RRx}) and find
\begin{equation}
\RR|x\rrangle=\QQ\left(|y_0\rrangle+c|0\rrangle\right)=\QQ|y_0\rrangle,
\end{equation}
since $\QQ|0\rrangle=0$.

In App.\ \ref{app:QR} we describe a simple numerical
algorithm for solving Eqs.\ (\ref{eq:hom_eq}) and (\ref{eq:Req}).
For very large dimensions of the involved matrices, it may
be necessary to invoke more advanced numerical methods to solve
these equations.\cite{Flindt2004} Numerically, the recursive scheme
is stable for very high orders of cumulants (up to order $\sim100$), which we have
tested on simple models. The results presented in this work have all
been obtained using standard numerical methods as the one described
in App.\ \ref{app:QR}.

\section{Asymptotics of high-order cumulants}
\label{sec:univosc}

Before illustrating our methods in terms of specific examples,
we discuss the asymptotic behavior of
high-order cumulants. As some of us have recently shown certain
ubiquitous features are expected for the high-order
cumulants.\cite{Flindt2009} In particular, the absolute values
of the high-order cumulants are expected to grow factorially  with the cumulant
order. Moreover, the high-order cumulants are predicted to oscillate as functions of
basically any parameter, as well as of the cumulant order. This
behavior was confirmed experimentally by measurements of the
high-order transient cumulants of electron transport through a
quantum dot.\cite{Flindt2009} In the experiment, the transient
cumulants indeed grew factorially  with the cumulant order and
oscillated as functions of time (before reaching the long-time
limit), in agreement with the general prediction. For completeness,
we repeat here the essentials of the theory underlying these
asymptotic properties of high-order cumulants.

The asymptotic behavior of high-order cumulants follows from
straightforward considerations. In the following we denote the CGF
by $S(\chi,\{\lambda\})$, where $\{\lambda\}$ represents the set of
all parameters needed to specify the system; whether the dynamics
is Markovian or non-Markovian is irrelevant. In general, we can
assume that the CGF has a number of singularities in the
complex-$i\chi$ plane at $i\chi=i\chi_j$, $j=1,2,3\ldots$, which can
be either poles or branch-points. Typically, the positions of the
singularities depend on  $\{\lambda\}$. Exceptions, where the CGF
has no singularities, do exist, e.\ g.\ the Poisson process, whose CGF
is given by an exponential function, but we exclude such cases in
the following.

Close to a singularity $i\chi\simeq i\chi_j$, we can write the CGF as
\begin{equation}\label{eq:singularity}
S(\chi,\{\lambda\})\simeq \frac{A_j}{(i\chi_j-i\chi)^{\mu_j}}
\end{equation}
for some $A_j$ and $\mu_j$, determined by the nature of the
singularity. For example, for a finite-order pole $\mu_j$ denotes the order of the pole, while $\mu_j=-1/2$ would correspond to the branch point of a square-root function. Logarithmic singularities can be treated on a similar footing with only slight modifications.\cite{Flindt2009} The derivatives with respect to the counting field are now
\begin{equation}
\frac{\partial^m S(\chi,\lambda)}{\partial(i\chi)^m}\simeq
\frac{A_jB_{m,\mu_j}}{(i\chi_j-i\chi)^{m+\mu_j}}
\end{equation}
with
\begin{equation}
B_{m,\mu_j}\equiv\mu_j(\mu_j+1)\cdots (\mu_j+m-1) \label{eq:Bmu}
\end{equation}
for $m\geq1$. As the order $m$ is increased this approximation
becomes better away from the singularity at
$\chi=\chi_j$ according to the Darboux theorem.\cite{Dingle1973,Berry2005,Flindt2009} For sufficiently high $m$,
the cumulants of the passed charge can thus be written
\begin{equation}
\begin{split}
\llangle
n^m\rrangle&=\left.\frac{\partial^mS(\chi,\lambda)}{\partial(i\chi)^m}\right|_{\chi\rightarrow0}\\
&\simeq\sum_j\frac{A_jB_{m,\mu_j}}{|i\chi_j|^{m+\mu_j}}e^{-i(m+\mu_j)\arg(i\chi_j)},
\end{split}
\label{eq:uniosc}
\end{equation}
where the sum runs over all singularities of the CGF. Here, we have written the singularities as
\begin{equation}
i\chi_j=|i\chi_j|e^{i\arg(i\chi_j)},
\end{equation}
where $|i\chi_j|$ is the modulus of the singularity $i\chi_j$ and
$\arg(i\chi_j)$ is the corresponding complex argument. In general,
the singularities $i\chi_j$ together with the factors $A_j$ come in complex conjugate pairs, ensuring that
the expression in Eq.\ (\ref{eq:uniosc}) is real.

From Eq.\ (\ref{eq:uniosc}) we deduce that the cumulants grow factorially
in magnitude with the order $m$ due to the factors $B_{m,\mu_j}$ given in Eq.\ (\ref{eq:Bmu}). We
also see that the high-order cumulants are determined primarily by
the singularities closest to zero. Contributions from other
singularities are suppressed with the relative
distance from zero and the order $m$, and can thus be neglected for
large $m$. Importantly, we observe
that the high-order cumulants become oscillatory functions of
\emph{any} parameter among $\{\lambda\}$ that changes
$\arg(i\chi_j)$ as well as of the cumulant order $m$ [see also Eq.\ (\ref{eq:twosingexp}) below]. We refer to these ubiquitous features, which should occur in a large class of
transport processes, as universal oscillations. For example, we expect oscillations of high-order cumulants for basically any transport process described by a GME, since the CGF for these systems typically have logarithmic singularities at finite times\cite{Flindt2009} or square-root branch points in the long-time limit.\cite{Bender1978} Factorial growth and oscillations as functions of various parameters can be found in several independent studies of high-order cumulants,\cite{Pilgram2003,Foerster2005,Foerster2007,Flindt2008,Urban2008,Khoruzhenko2009,Prolhac2009,Golubev2010,Hassler2010} as well as in the recent experiment described in Ref.\ \onlinecite{Flindt2009}, demonstrating the generality of the phenomenon. Similar observations and discussions can also be found in quantum optics\cite{Dodonov1994} and high-energy physics,\cite{Dremin1994,Bhalerao2003,Bhalerao2004} further confirming the prediction. We note that in the long-time limit, the positions of the dominating singularities are
no longer time dependent,\cite{Flindt2009} and the cumulants cease
to oscillate as functions of time. Instead, the cumulants of the
passed charge become linear in time, as previously discussed in Sec.\
\ref{sec:zerofreqcumu}.

A simple (and common) situation arises if only two complex conjugate singularities, $|i\chi_0|e^{i\arg{i\chi_0}}$ and $|i\chi_0|e^{-i\arg{i\chi_0}}$, are closest to zero.
In that case, Equation (\ref{eq:uniosc}) immediately yields
\begin{equation}\label{eq:twosingexp}
    \llangle n^m\rrangle\simeq \!\frac{2|A_0|
    B_{m,\mu_0}}{|i\chi_0|^{m+\mu_0}}\cos\!\left[(m\!+\!\mu_0)\arg i\chi_0\!-\!\arg
    A_0\right].
\end{equation}
Using this expression we can determine the positions of the dominating singularities from numerical calculations of the high-order cumulants as we shall demonstrate in the second example considered in Sec.\ \ref{sec:Markexamples}. We note that while the factorial growth and the oscillations are system independent, other features, for example the frequency of the oscillations, are determined by the particular details of the system under consideration.

Finally, we mention the Perron-Frobenius theorem regarding stochastic matrices\cite{Baiesi2009,Demboo1998} which implies that the CGFs considered in Sec.\ \ref{sec:Markexamples} must be analytical functions at least in a strip along the real axis in the
complex-$i\chi$ plane.  This has important consequences especially for the nature of the high-order
cumulants which rests heavily on the analytical properties of the CGF. We illustrate this statement in both examples in Sec.\ \ref{sec:Markexamples}.

\section{Markovian Systems}
\label{sec:Markexamples}

\subsection{Electron bunching in a two-level quantum dot}
\label{sec:Belzig}

In our first example we study electron bunching in transport
through a two-level quantum dot as described by Belzig in
Ref.~\onlinecite{Belzig2005}. Due to the relatively simple
analytical structure of the model, it is possible to
illustrate the concepts of universal oscillations introduced
above. The model allows us to test the accuracy of our numerical
calculations of high-order cumulants against analytic
expressions.

We start by summarizing the setup in Belzig's model. Consider a
single quantum dot with two single-particle levels coupled to
voltage-biased source and drain electrodes. The two levels serve as
parallel transport channels. Due to strong Coulomb interactions on
the quantum dot only one of the levels can be occupied at a time.
The system exhibits super-Poissonian bunching transport in cases
where both levels are coupled by the same rate $\Gamma_L$ to the,
say, left lead, whose Fermi level is kept well above both levels,
while the couplings to the other lead are markedly different, such
that one level is coupled to the right lead by the rate $\Gamma_R\ll
\Gamma_L$ and the other by $x\Gamma_R$ with $x\ll 1$. This situation
can arise for example, if the two levels are situated above and
slightly below, respectively, the Fermi level of the right lead at a
finite electron temperature.

This particular configuration leads to bunching of electrons in
the transport due to the existence of the blocking state: if the
dot is empty there is equal probability for either of the two
levels to be filled. Current runs easily through the first level,
while the other level effectively is blocked, or more precisely,
the transport through the level is limited by the very small right
rate $x\Gamma_R$, constituting a bottleneck. The transport thus proceeds in
bunches of electrons passing intermittently through the first
level separated by quiet periods of blocked transport when
the other level is occupied. This bunching effects leads to super-Poissonian noise with a
Fano factor above unity. For more detailed discussions of the model
as well as its generalizations to many levels, the reader is
referred to Ref.~\onlinecite{Belzig2005}.

The counting statistics of the system can be obtained from a
Markovian rate equation for the probability vector
$\hat{p}=(p_0,p_+,p_-)^T$, containing the ($n$-resolved)
probabilities $p_{0,-,+}$ for the quantum dot to be empty, or the
first ($+$, non-blocking) or second ($-$, blocking) level being
occupied, respectively. The corresponding $\chi$-dependent rate
matrix reads
\begin{equation}
\label{eq:Wolfgangmodel}
    \W (\chi)
    =\begin{pmatrix}
    -2-\Gamma(1-x) & \Gamma e^{i\chi} & x\Gamma e^{i\chi}\\
    1              & -\Gamma          & 0\\
    1+e^{-i\chi}\Gamma(1-x) & 0       & -x\Gamma
    \end{pmatrix}
\end{equation}
Here, we have rescaled the time and set $\Gamma_L\equiv 1$ while
renaming $\Gamma_R\equiv\Gamma$ in order to simplify the analytic
results in the following. We have also made a minor modification
of the model in Ref.\ \onlinecite{Belzig2005} by including the
back-flow into the blocking level from the right lead. This
modification, however, changes only slightly the detailed
quantitative results, while leaving the main qualitative features
identical in the limit of interest $x,\Gamma\ll 1$.

Since the model involves only three states, the CGF can be found
analytically in the long-time limit. The full expression is too
lengthy to be presented here, but in the limit $x, \Gamma \ll 1$,
it reduces to the result by Belzig\cite{Belzig2005} (also for our
slightly modified model; note, however, the opposite sign
convention for the CGF in Ref.~\onlinecite{Belzig2005})
\begin{equation}
    S(\chi,t)\rightarrow 2\Gamma xt\frac{e^{i\chi}-1}{2-e^{i\chi}}.
\label{eq:Belzigsinglepole}
\end{equation}
Clearly, the CGF has simple poles at
\begin{equation}
i\chi_j=\ln 2+j2\pi i,\,\,\, j=\ldots,-1,0,1,\ldots
\end{equation}
with the pole $i\chi_0=\ln 2$ being closest to 0. However, according to the Perron-Frobenius theorem mentioned in Sec.\ \ref{sec:univosc} the CGF cannot have singularities on the real $i\chi$-axis.

In order to illustrate this point, we consider the expected behavior
of the high-order cumulants based on the CGF above. Close to the
singularity $i\chi_0$, we approximate the CGF by the first non-zero term of the Laurent series
\begin{equation}
    S(\chi,t)\simeq \frac{\Gamma x t}{i\chi_0-i\chi}.
\end{equation}
This corresponds to Eq.\ (\ref{eq:singularity}) with $A_0=\Gamma
x t$ and $\mu_0=1$. From Eq.\ (\ref{eq:uniosc}) we then obtain a
simple asymptotic expression for the high-order cumulants reading
\begin{equation}
\llangle I^m\rrangle_{\!1s}/\Gamma x=\llangle n^m\rrangle_{\!1s}/\Gamma x t \simeq m!/(\ln 2)^{m+1}.
\label{eq:Belzig1pApprox}
\end{equation}
Here, the subscript $_{1s}$ indicates that the expression has been
obtained using the approximate CGF in Eq.\ (\ref{eq:Belzigsinglepole}) with
only a single singularity closest to zero. In Table
\ref{table:comparison} we compare the asymptotic expression with
results for the first six cumulants obtained by direct
differentiation of the CGF in Eq.\ (\ref{eq:Belzigsinglepole}).
The asymptotic results are very close to the exact derivatives of
the approximate CGF. Despite the good agreement with the
approximate results, the asymptotic expression in Eq.\
(\ref{eq:Belzig1pApprox}) does not reproduce our numerically exact
results, also shown in the table, obtained using our recursive scheme. In particular for high orders,
the asymptotic expression starts to deviate significantly from the
numerically exact results.

\begin{table}
\begin{tabular}{|l|c|c|c|c|c|c|}
\hline
\hline
$\llangle I^m\rrangle/\Gamma x$ & $m=1$ & 2 & 3 & 4 & 5 & 6 \\
\hline
Single-pole approx. & 2.000 & 6.000 & 26.00 & 150.0 & 1082 & 9366\\
\hline
Single-pole asympt. & 2.081 & 6.006 & 25.99 & 150.0 & 1082 & 9366\\
\hline
Numerics & 1.978 & 5.880 & 25.18 & 143.3 & 1017 & 8644\\
\hline
\hline
\end{tabular}
\caption{Normalized zero-frequency current cumulants for transport
through a two-level quantum dot. Single-pole approximation results
have been obtained by direct differentiation of the CGF in Eq.\
(\ref{eq:Belzigsinglepole}) or its asymptotic expression Eq.\
(\ref{eq:Belzig1pApprox}), respectively. The numerically exact
results have been obtained using our recursive scheme and the rate
matrix in Eq.\ (\ref{eq:Wolfgangmodel}) with $x=0.001$ and
$\Gamma=0.01$.} \label{table:comparison}
\end{table}

\begin{figure*}
\begin{center}
\includegraphics[width=0.90\textwidth, trim = 0 0 0 0, clip]{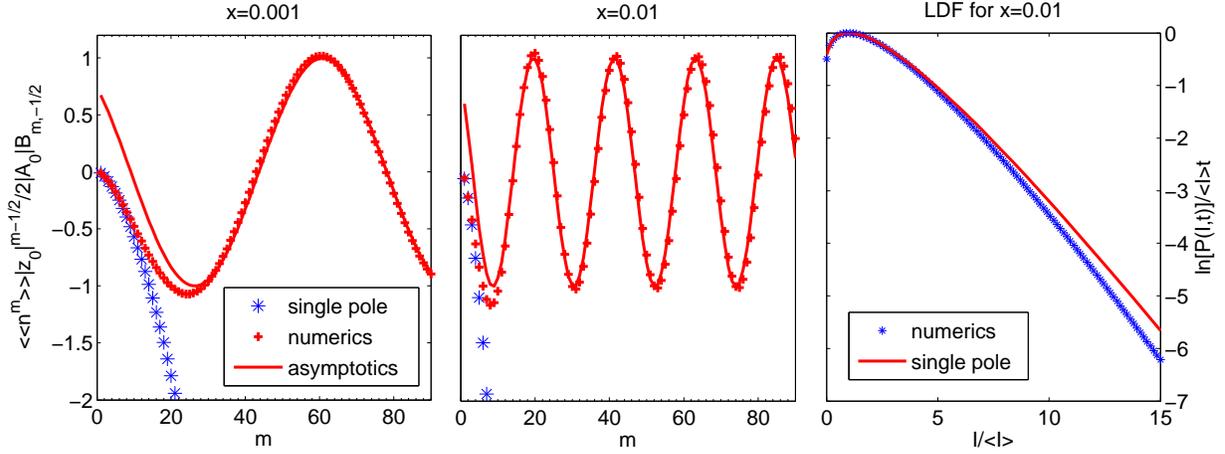}
\caption{High-order cumulants and large deviation function for bunching
transport through a two-level quantum dot. Left and central panels show comparisons between exact
numerics and the single
pole approximation stemming from Eq.\ (\ref{eq:Belzigsinglepole}) for two different
values of $x=0.001,\,0.01$ and $\Gamma=0.01$. The asymptotic expression in Eq.\ (\ref{eq:twosingexp}) based on a pair of complex conjugate singularities is shown with full lines. Notice that $B_{m,-1/2}<0$. The right panel shows a comparison of the large deviation function (LDF) obtained from exact numerics and the single pole approximation in Eq.\ (\ref{eq:Belzig1pLDF}), respectively.} \label{fig:bunchingorder}
\end{center}
\end{figure*}

As anticipated above, these deviations can be traced back to the
expression in Eq.\ (\ref{eq:Belzigsinglepole}), that we obtained
in the limit $x, \Gamma \ll 1$. In order to proceed from here, we return to the full expression
for the CGF in the long-time limit (not shown). A careful analysis
reveals that, in fact, there is a pair of complex conjugate
singularities closest to zero, and not just a single pole. The two
singularities, denoted as $i\tilde{\chi}_0$ and
$(i\tilde{\chi}_0)^*$, correspond to branch points of a
square-root, and for small $x\ll1$ the position of the branch
point $i\tilde{\chi}_0$ is
\begin{equation}\label{eq:Belzigpoleposition}
    i\tilde{\chi}_0 = \ln(2+\Gamma)-2 x\frac{4 + \Gamma (6 + \Gamma)}{(2 +
    \Gamma)^2}+ 4i \sqrt{x}\frac{1 + \Gamma}{2 + \Gamma}.
\end{equation}
Clearly, for small $x,\Gamma\ll 1$ the branch points are close to
the position of the single pole $i\chi_0=\ln 2$. However, for any
finite $x$, the two branch points have small, but finite,
imaginary parts thus complying with the Perron-Frobenius theorem. The singularity structure around the branch point $i\tilde{\chi}_0$
is characterized by Eq.~\eqref{eq:singularity} with $\mu_0=-1/2$ and
$A_0\approx \Gamma t\sqrt[4]{x}e^{i\pi/4}$, and we can then use the asymptotic expressions in Eq.\ (\ref{eq:twosingexp}) for the high-order cumulants.
In the left and central panels of Fig.~\ref{fig:bunchingorder} we compare this expression, and the single-pole
approximation in Eq.\ (\ref{eq:Belzig1pApprox}), with numerically exact results obtained using our
recursive scheme for $\Gamma=0.01$ and two different values of $x=0.001,\,0.01$.

Figure~\ref{fig:bunchingorder} shows several important features.
Firstly, the (scaled) high-order cumulants indeed behave in an
oscillatory manner as function of the cumulant order $m$, which
coincides with the cosine part of  Eq.\
(\ref{eq:twosingexp}). Obviously, for smaller $x=0.001$ the period
of the oscillations, determined by $\arg i\tilde{\chi}_0$, is longer
in accordance with Eq.\ (\ref{eq:twosingexp}). Furthermore,
for the small value of $x=0.001$, the asymptotic form of the
high-order cumulants is reached around $m\simeq 30$, while the
single-pole approximation agrees well for lower orders, $m\lesssim
10$. For the higher value of $x=0.01$, significant deviations from
the single-pole behavior begin already for the fourth cumulant,
while the asymptotic oscillatory form holds from around $m=12$.
Notice the importance of the exact analytical knowledge of the
singularities
--- even though $x=0.01\ll 1$ (together with $\Gamma=0.01\ll 1$) may
seem a very small number justifying the usage of the single pole
approximation, we see from Eq.\ (\ref{eq:Belzigpoleposition}) that
the imaginary part of the pole and its argument scale like
$\sqrt{x}=0.1$, thus invalidating the single-pole approximation
far earlier than expected from a linear-in-$x$ scaling assumption.

A complementary view on the charge transport statistics is provided by the large deviation function (LDF),\cite{Touchette2009} which quantifies deviations of measurable
currents from the average value. The LDF is obtained from the probability
distribution
\begin{equation}
P(n,t)=\frac{1}{2\pi}\int_{-\pi}^{\pi}d\chi e^{S(\chi,t)-in\chi}
\end{equation}
and is defined as the long-time limit of $\ln[P(I,t)]/t$, where $I\equiv n/t$ is the current. For long times, we have $S(\chi,t)\rightarrow \lambda_0(\chi) t$ and the integral can be evaluated in the saddle-point approximation with the saddle-point $\chi=\chi_0$ given by the solution to the saddle-point equation
\begin{equation}
\lambda_0'(\chi_0)=iI, \label{eq:saddlepointeq}
\end{equation}
The saddle-point equation implies a parametric dependence of the saddle-point $\chi_0=\chi_0(I)$ on the current $I$. Using the saddle-point approximation, the LDF becomes
\begin{equation}
\frac{\ln [P(I,t)]}{t}\rightarrow \lambda_0(\chi_0)-iI\chi_0. \label{eq:largedeviation}
\end{equation}
We first solve the saddle-point equation for the approximate CGF in Eq.\ (\ref{eq:Belzigsinglepole}) and find
\begin{equation}
\frac{\ln [P_{1s}(I,t)]}{\langle I\rangle t}\rightarrow\frac{\sqrt{1+8 \kappa}-3}{4}
-\kappa\log\left[\frac{16\kappa}{(1+\sqrt{1+8\kappa})^2}\right], \label{eq:Belzig1pLDF}
\end{equation}
where $\kappa\equiv I/\langle I\rangle$ and the subscript $_{1s}$ again reminds us that the expression has been
obtained using the approximate CGF with only a single singularity closest to zero. Obviously, the current must be positive ($\kappa>0$), since transport is unidirectional.

Also for the LDF, we can compare the analytic approximation with numerical exact results. To this end, we need to solve
the saddle-point equation taking as starting point the kernel in Eq.\ (\ref{eq:Wolfgangmodel}).
The derivative of the eigenvalue $\lambda_0(\chi)$ is now calculated using the
Hellman-Feynman theorem, writing
\begin{equation}
\begin{split}
\lambda_0'(\chi)&=\frac{\partial}{\partial\chi}\llangle \tilde{0}(\chi)|\W(\chi)|0(\chi)\rrangle \\
&=\llangle \tilde{0}(\chi)|\W'(\chi)|0(\chi)\rrangle,
\end{split}
\label{eq:Hellman-Feynman}
\end{equation}
where $\llangle\tilde{0}(\chi)|$ and $|0(\chi)\rrangle$ are left and
right eigenvectors of $\W(\chi)$, respectively, corresponding to the
eigenvalue $\lambda_0(\chi)$, and
$\llangle\tilde{0}(\chi)|0(\chi)\rrangle=1$. For a given value of $\chi$ we calculate numerically the left and right eigenvectors $\llangle \tilde{0}(\chi)|$ and $|0(\chi)\rrangle$ and find $\lambda_0'(\chi)$ using the expression for the derivative in Eq.\ (\ref{eq:Hellman-Feynman}). With this procedure we search numerically for the value of $\chi=\chi_0$ that solves Eq.\ (\ref{eq:saddlepointeq}) for a given value of $I$, and with the solution $\chi_0$ we evaluate the LDF using Eq.\ (\ref{eq:largedeviation}). We find that $\chi_0$ is purely imaginary.\cite{Bagrets2003} We note that, in principle, the existence of a saddle point solution is not guaranteed in the whole range of currents, and there are examples,\cite{Bagrets2006,Visco2006} where the behavior of the LDF changes abruptly at finite values of $I$ due to singularities of the CGF on the real $i\chi$-axis. In our case, however, the Perron-Frobenius theorem ensures that the CGF is analytic on the real $i\chi$-axis and the LDF is smooth as function of $I$. In the right panel of Fig.\ \ref{fig:bunchingorder} we show a comparison between exact numerics and the analytic result (\ref{eq:Belzig1pLDF}) for the large deviation function in the single pole approximation. Around the mean value $I\simeq \langle I\rangle$ the analytic result agrees well with numerics. However, in the tails of the distribution a clear disagreement between the analytic approximation and numerics is visible. The disagreement reflects the deviations for the cumulants seen in the central panel of Fig.\ \ref{fig:bunchingorder}. We remark that measurements of the LDF recently have become accessible in experiments on real-time electron counting.\cite{Fricke2010}

The discussion in this subsection illustrates the need for careful considerations when
manipulating CGFs analytically. Concerning cumulants, we deal with
two opposite and non-commutative orders of limits: for a fixed order
of cumulants, a limiting procedure with changing parameters
converges to the approximate form given by the appropriate limit of
the CGF, such as the single-pole approximation in Eq.\
(\ref{eq:Belzigsinglepole}) in our case. However, the convergence of
the CGF is not uniform in $\chi$ due to potential singularities and
thus for fixed parameters, high order cumulants generically take on
the universal oscillatory form discussed above. One should thus be careful when using limiting forms of a
CGF to extract cumulants of arbitrary orders. In general, the
low-order cumulants  follow the predicted pattern reasonably well,
but at some point significant deviations appear and the universal
oscillatory behavior should emerge. The order at which this crossover
occurs depends on details of the analytical structure of the CGF and
may be hard to predict. As we have shown explicitly, deviations of the cumulants from exact results are also clearly visible in the large deviation function.

\subsection{Transport through a vibrating molecule}
\label{subsec:vibration}

\begin{figure}
\begin{center}
\includegraphics[width=0.45\textwidth]{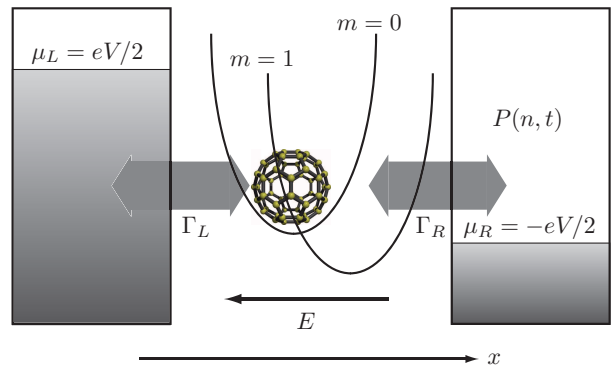}
\caption{(color online). Transport through vibrating molecule. The
molecule is coupled to the left (right) lead with coupling
$\Gamma_L$ ($\Gamma_R$). The bias difference $eV=\mu_L-\mu_R$ drives
single electrons through the molecule. The system is operated in the
Coulomb blockade regime, where only $m=0$ or $m=1$ additional
electrons are allowed on the molecule. As an electron tunnels onto
the molecule, the equilibrium position of the molecule is shifted
due to the electric field $E$. The two harmonic
potentials corresponding to $m=0,1$ are shown. The damping rate of the vibrating molecule is denoted as $K$.} \label{fig:vibmole}
\end{center}
\end{figure}

In our next example, we consider a model of charge transport through a
molecule coupled to quantized vibrations.\cite{McCarthy2003,Millis2004,Boese2001,Flindt2005,Koch2005,Koch2005b,Haupt2006}
In the regime of weak coupling to the electronic leads, electron tunneling can be described using Fermi's golden rule rates for transitions between different vibrational and charge occupation states. For strong electron-phonon coupling, the large shift of the oscillator equilibrium position due to an electron tunneling onto the molecule suppresses the (Franck-Condon) overlap between the initial and final vibrational state for low-lying oscillator states. This leads to surpressed tunnel rates at low
bias-voltages, so-called Franck-Condon blockade. For larger bias-voltages, higher-excited oscillator
states become available, and the system can escape the blockade
regime. For weak oscillator dampings, several electrons can be
transferred through the molecule, once the blockade is lifted, until
a charge transfer event eventually leaves the oscillator in the
ground state and the current is suppressed again. Such dynamical
Franck-Condon blockade processes have been predicted to lead to very
large enhancements of the zero-frequency noise.\cite{Koch2005} Recently, Frank-Condon blockade was observed in experiments on suspended carbon nanotube quantum dots.\cite{Leturcq2009}

The system considered in the following is depicted in Fig.\ \ref{fig:vibmole}. Here we follow to
a large extent the description of the model given in Refs.\ \onlinecite{McCarthy2003}, \onlinecite{Flindt2005}. The Hamiltonian of the system and the detailed derivation of the resulting Markovian GME are given in App.\ \ref{app:vibmol}, where the various parameters of the model are also defined.
 Due to the large number of
oscillator states, there is little hope for obtaining a closed-form
expression for the CGF that would allow for any analytic
manipulations. Instead, as we shall see, the numerically
evaluated high-order cumulants can be used to extract the precise
location of the dominating singularities of the CGF. We concentrate in the following on the unequilibrated oscillator
regime, where the damping rate of the oscillator is much smaller than the electron tunneling rates, $K\ll\Gamma_{L/R}$. As explained above, the combination of
strong electron-phonon coupling and weak oscillator damping leads to
dynamical Franck-Condon blockade, resulting in a large enhancement
of the current noise as demonstrated in Ref.\ \onlinecite{Koch2005}
using Monte-Carlo simulations. In Ref.\ \onlinecite{Koch2005b} the
analysis was extended to the full distribution of the transferred
charge and an analytic approximation for the CGF was presented based
on an avalanche-type of transport, where ``quiet'' periods of
transport are interrupted by a sequel of self-similar charge
avalanches. The analytic result for the CGF was shown to agree very
well with Monte-Carlo simulations of the probability distribution
$P(n,t)$. However, similarly to the previous
example, the approximate CGF has a single, simple pole on the
real-$i\chi$ axis, violating the required properties of the CGF,
mentioned at the end of Sec.\ \ref{sec:univosc}, thus making it unsuited for
predictions of the high-order cumulants. In particular, within this
approximation, the high-order cumulants would not oscillate, which contradicts our numerical findings.

Oscillations of the high-order cumulants with system parameters must be due to singularities located away from the real-$i\chi$. In the following,  we assume that the CGF has a pair of complex-conjugate
singularities, $i\chi_0=|i\chi_0|e^{i\arg i\chi_0}$ and $|i\chi_0|e^{-i\arg i\chi_0}$, closest
to zero. As we will now show, the positions of these singularities can be found from our numerical calculations of high-order cumulants. To this end, we define
\begin{equation}
a_0=A_0/(i\chi_0)^{\mu_0}
\label{eq:a0}
\end{equation}
and rewrite Eq.\ (\ref{eq:twosingexp}) as
\begin{equation}
    \llangle n^m\rrangle\simeq \!\frac{2|a_0|
    B_{m,\mu_0}}{|i\chi_0|^{m}}\cos\!\left[m\arg i\chi_0\!-\!\arg
    a_0\right].
    \label{eq:twosingexpre}
\end{equation}
Following the ideas of Ref.\ \onlinecite{Zamastil2005}, Sec.\ 4, we find for the ratios of two successive cumulants
\begin{equation}
\begin{split}
\frac{\llangle n^{m-1}\rrangle}{\llangle n^m\rrangle}\frac{m+\mu_0-1}{|i\chi_0|}&=\cos\!\left[\arg i\chi_0\right]\\
&+\sin\!\left[\arg i\chi_0\right]\tan\!\left[m\arg i\chi_0-\!\arg a_0\right]
\end{split}
\end{equation}
and
\begin{equation}
\begin{split}
\frac{\llangle n^{m+1}\rrangle}{\llangle n^m\rrangle}\frac{|i\chi_0|}{m+\mu_0}&=\cos\!\left[\arg i\chi_0\right]\\
&-\sin\!\left[\arg i\chi_0\right]\tan\!\left[m\arg i\chi_0-\!\arg a_0\right].
\end{split}
\end{equation}
Adding the two left and right hand sides, respectively, and rearranging, we obtain the equation
\begin{equation}
\begin{split}
2(m+\mu_0)\llangle n^{m}\rrangle |i\chi_0|\cos\!\left[\arg i\chi_0\right]-\llangle n^{m+1}\rrangle|i\chi_0|^2=\\
\llangle n^{m-1}\rrangle(m+\mu_0-1)(m+\mu_0).
\end{split}
\end{equation}
Using the substitution $m\rightarrow m+1$, we obtain an additional equation and thus arrive at a linear system of two equations which we solve for
$|i\chi_0|\cos\!\left[\arg i\chi_0\right]$ and $|i\chi_0|^2$ and thereby find $i\chi_0$. The method takes as input $\llangle n^{m-1}\rrangle$, $\llangle n^{m}\rrangle$, $\llangle n^{m+1}\rrangle$, and $\llangle n^{m+2}\rrangle$, and the accuracy is expected to improve with increasing cumulant order $m$.\cite{Zamastil2005}

\begin{figure}
\begin{center}
\includegraphics[width=0.45\textwidth, trim = 0 0 0 0, clip]{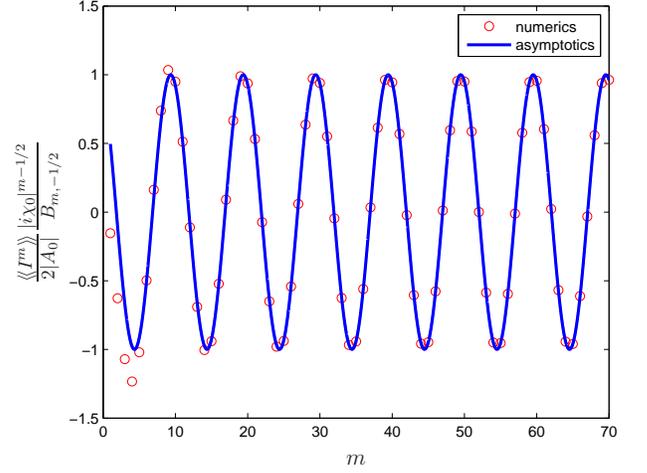}
\caption{(color online). High-order (normalized) cumulants for
unequilibrated molecule. Numerically exact results are shown
together with the asymptotics  described by Eq.\
(\ref{eq:twosingexp}). Parameters entering Eq.\ (\ref{eq:twosingexp}) are
$\mu_0=-1/2$, $A_0=1.4810\times 10^{-7}e^{-i0.7378}$, and
$i\chi_0=0.0113 e^{i0.6262}$. System parameters (defined in App.\ \ref{app:vibmol}) are given in units of the natural oscillator frequency (with
$e,\hbar,k_B=1$) $V=3\omega_0$, $\Gamma=\Gamma_L=\Gamma_R=0.001\omega_0$,
$T=0.05\omega_0$, $K=10^{-10}\omega_0$, $\varepsilon=16\omega_0$, $c_1=4$, $c_2=0$. In the numerical calculations we have used $N=15$ oscillator states.} \label{fig:C60res2}
\end{center}
\end{figure}

Having determined $i\chi_0$, we find $a_0$ in a similar spirit by rewriting Eq.\ (\ref{eq:twosingexpre}) as
\begin{equation}
\begin{split}
    \llangle n^m\rrangle\simeq &2B_{m,\mu_0}\left[\mathrm{Re}\{(i\chi_0)^{-m}\}\mathrm{Re}\{a_0\}\right.\\
    &\left.
    -\mathrm{Im}\{(i\chi_0)^{-m}\}\mathrm{Im}\{a_0\}\right].
\end{split}
\end{equation}
Again, we obtain via the substitution $m\rightarrow m+1$ a linear system of two equations that we solve for $\mathrm{Re}\{a_0\}$ and $\mathrm{Im}\{a_0\}$ and thus find $a_0=\mathrm{Re}\{a_0\}+i\mathrm{Im}\{a_0\}$. Finally, we determine $A_0$ from Eq.\ (\ref{eq:a0}). More advance methods for extracting the positions of singularities are available,\cite{Zamastil2005} but they require solutions of non-linear equations and will not be considered here.

In order to extract $i\chi_0$ and $A_0$ from the high-order cumulants, we need to know the nature of the singularities and hence $\mu_0$. Typically, the singularities are square-root branch points (see Ref. \onlinecite{Bender1978}, Sec.\ 7.5) and we thus take $\mu_0=-1/2$. In Fig.\ \ref{fig:C60res2} we show numerical results for the
(normalized) cumulants as function of the order $m$ together with
the asymptotic expression Eq.\ (\ref{eq:twosingexp}) for the high-order cumulants with $i\chi_0$ and $A_0$ found using the method described above. The asymptotic expression shows excellent agreement with the numerically exact results. For $m\gtrsim 5$, we see trigonometric oscillations whose frequency is determined by $\arg{i\chi_0}$. We note that a good agreement between our numerical results and the asymptotic expression could only be obtained with $\mu_0=-1/2$, thus confirming that the singularities stem from square-root branch points.

\begin{figure}
\begin{center}
\includegraphics[width=0.45\textwidth, trim = 0 0 0 0, clip]{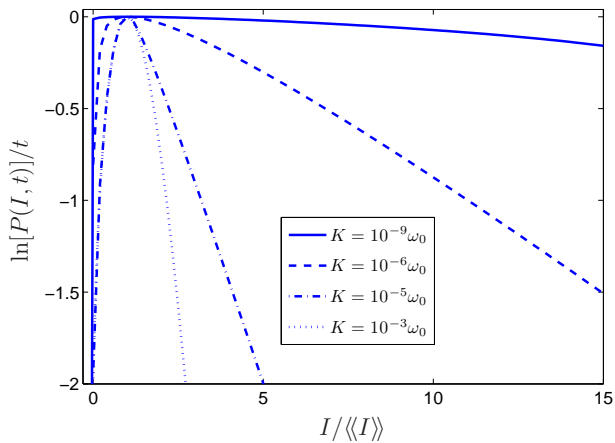}
\caption{(color online). Large deviation function for the vibrating
molecule. Results are shown for different values of the damping
$K$, going from the unequilibrated regime, at low $K\ll\Gamma$, to
$K\sim\Gamma$, where the molecule equilibrates between each
tunneling event. For the unequilibrated case, a much larger range
of currents are probable, compared to transport through the
equilibrated molecule. System parameters (defined in App.\ \ref{app:vibmol}) are given in units of the natural oscillator frequency (with
$e,\hbar,k_B=1$) $V=3\omega_0$, $\Gamma_L=\Gamma_R=0.001\omega_0$,
$T=0.05\omega_0$, $\varepsilon=16\omega_0$, $c_1=4$, $c_2=0$.} \label{fig:C60res3}
\end{center}
\end{figure}

The large deviation function can also be evaluated numerically using the method described in the previous subsection.
In Fig.\ \ref{fig:C60res3} we show numerical results for the large
deviation function with different values of the damping $K$. For
large dampings, the oscillator is essentially equilibrated and the
measurable currents are closely centered around the mean current
$\llangle I\rrangle$. As the damping is lowered, we approach the
unequilibrated regime, where the transport statistics is dominated
by avalanche transport with a corresponding large zero-frequency
noise. Accordingly, the large deviation function is considerably
broadened and a much wider range of currents
become measurable.

\section{Non-Markovian systems}
\label{sec:nonMarkov}
\subsection{Dissipative double quantum dot}
\label{subsec:DQD}

In the previous two examples, we focused on the asymptotic behavior of
the high-order cumulants for two Markovian systems. We now turn
our attention to a model for which a weak coupling prescription
does not suffice and non-Markovian effects become significant. We
focus here on the influence of memory effects on the first few
cumulants, while referring the reader to Ref.\
\onlinecite{Flindt2008} for a discussion of the high-order
cumulants for the non-Markovian system presented in this example.

We consider a model of charge transport through a double
quantum dot (DQD) coupled to a heat bath which causes dephasing and
relaxation. Such systems were studied experimentally in Refs.\ \onlinecite{Fujisawa1998,Barthold2006}. The counting statistics in the transition between
coherent and sequential tunneling through DQDs has
been studied theoretically by Kie{\ss}lich and
co-workers.\cite{Kiesslich2006} In their work, decoherence was
described using either a charge detector model or via
phenomenological voltage probes.\cite{Blanter2000} More elaborate descriptions of
decoherence caused by a weakly coupled heat bath were given in Refs.\
\onlinecite{Kiesslich2007,Sanchez2008} and shown to agree well with experiments.

Here, we take these ideas further and go beyond the perturbative
treatment of the heat bath. This situation has previously been investigated by
Aguado and Brandes using a polaron transformation, assuming weak
coupling to the electronic leads in the high-bias
limit.\cite{Brandes1999,Aguado2004a,Aguado2004b} In the following, we apply an
alternative non-perturbative scheme for the coupling to the heat
bath, enabling us to fully include broadening due to the electronic leads.
Within this approach, we can study the cross-over between weak and
strong couplings to the heat bath and evaluate the effects of strong
decoherence on the charge transport statistics. In particular, we
show that only in the limit of weak coupling and high temperatures,
the dephasing caused by the heat bath can be accounted for by a
charge detector model with a single effective dephasing rate.

\begin{figure}
\begin{center}
\includegraphics[width=0.45\textwidth]{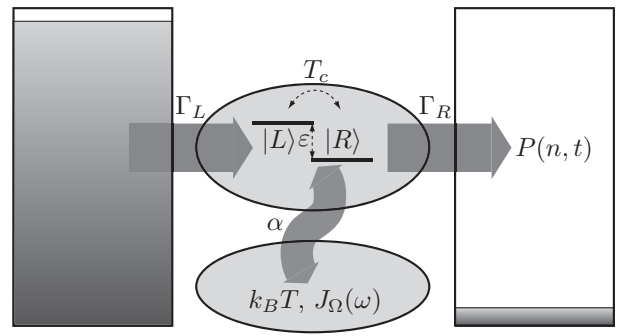}
\caption{Dissipative double quantum dot. The Coulomb blockaded
double quantum dot consists of the left and right levels $|L\rangle$
and $|R\rangle$, coherently coupled with tunnel coupling $T_c$ and
dealigned by $\varepsilon$.  A large bias across the system drives
electrons through the double quantum dot from the left lead with
rate $\Gamma_L$ to the right lead with rate $\Gamma_R$. The system
is coupled with dissipation strength $\alpha$ to a heat bath
at temperature $T$ and with Ohmic spectral function $J_{\Omega}(\omega)$. The
probability distribution of the number of transferred charges $n$ is
denoted $P(n,t)$.} \label{fig:DQDsetup}
\end{center}
\end{figure}

The model of charge transport through a Coulomb blockaded
DQD\cite{Aguado2004a,Aguado2004b} is
illustrated in Fig.\ \ref{fig:DQDsetup}. The DQD is coupled
to source and drain electrodes, while dissipation is provided by an
external heat bath. The DQD is operated in the Coulomb blockade regime close to a charge
degeneracy point, where only a single additional electron is allowed
on the double dot. Again, we consider for simplicity spinless electrons. The Hamiltonian of the double dot
can be written
\begin{equation}
\hatH_S=\epsilon_0|0\rlangle
0|+\frac{\varepsilon}{2}\hat{s}_z+T_c\hat{s}_x,
\end{equation}
where the pseudo-spin operators are
\begin{equation}
\hat{s}_z\equiv |L\rlangle L|-|R\rlangle R|
\end{equation}
and
\begin{equation}
\hat{s}_x\equiv |L\rlangle R|+|R\rlangle L|,
\end{equation}
respectively. Here, the two quantum dot levels $|L\rangle$ and
$|R\rangle$ are dealigned by $\varepsilon$ and their tunnel coupling
is $T_c$. The energy of the `empty' state $|0\rangle$ is
$\epsilon_0$. The pseudo-spin interacts with an external heat bath
consisting of harmonic oscillators,
\begin{equation}
\hatH_B=\sum_j\hbar\omega_j\hat{a}_j^{\dagger}\hat{a}_j,
\end{equation}
whose positions are coupled to the $z$-component of the pseudo-spin,
adding the term $\hat{V}_B\hat{s}_z$ to the full Hamiltonian with
\begin{equation}
\hat{V}_B= \sum_j\frac{g_j}{2}(\hat{a}_j^{\dagger}+\hat{a}_{j}).
\end{equation}
Finally, the spin-boson system is tunnel-coupled to left ($L$) and
right ($R$) leads via the tunnel-Hamiltonian
\begin{equation}
\hatH_T=\sum_{k_{\alpha},\alpha=L,R}(t_{k_{\alpha}}\hat{c}^{\dagger}_{k_\alpha}|0\rlangle
\alpha|+\mathrm{h.c.}),
\end{equation}
with both leads described as non-interacting fermions, i.e.,
\begin{equation}
\hatH_{\alpha}=\sum_{k_\alpha}\varepsilon_{k_\alpha}\hat{c}^{\dagger}_{k_\alpha}\hat{c}_{k_\alpha},\,\,\,
\alpha=L,R,
\end{equation}
kept at chemical potentials $\mu_{\alpha}$, $\alpha=L,R$, and temperature $T$. The full
Hamiltonian then reads
\begin{equation}
\hatH=\hatH_S+\hatH_T+\hatH_L+\hatH_R+\hatH_B+\hat{V}_B\hat{s}_z.
\label{eq:fullHamiltonian}
\end{equation}
As previously pointed out,\cite{Aguado2004a,Aguado2004b} the model
can be mapped onto that of transport through a superconducting single
electron transistor, when the charging energy is much larger than
the Josephson energy. Throughout this example we take $\hbar=k_B=e=1$.

As explained in App.\ \ref{app:doubledot}, transport through the
double dot can be described using a non-Markovian equation of motion
of the form in Eq.\ (\ref{eq:GME}) for the three electronic
occupations of the double dot collected in the vector
$\hatrho=(\rho_{0},\rho_{L},\rho_{R})^T$. The occupation
probabilities of the empty, left, and right states, are denoted
$\rho_{0}$, $\rho_{L}$, and $\rho_{R}$, respectively. The
corresponding memory kernel in Laplace space reads
\begin{equation}
\W(\chi,z)=
\begin{pmatrix}
  -\Gamma_L & 0     & \Gamma_Re^{i\chi}\\
  \Gamma_L  & -\Gamma_{B}^{(+)}(z)     & \Gamma_{B}^{(-)}(z)         \\
  0         & \Gamma_{B}^{(+)}(z)     & -\Gamma_{B}^{(-)}(z)-\Gamma_R \\
\end{pmatrix}.
\label{eq:kernel}
\end{equation}
We note that the kernel with $\chi=0$ has a single zero-eigenvalue $\lambda_0(0,z)=0$ for all $z$, in agreement with Eq.\ (\ref{eq:lambda0}). The kernel has been derived under the assumption that
the symmetrically applied
bias $eV=|\mu_L-\mu_R|$ between the electronic leads is much larger than
the tunneling rates to the leads and the temperature $T$. The tunneling rates are defined as
\begin{equation}
\Gamma_{\alpha}(\epsilon)=2\pi\sum_{k}|t_{k_{\alpha}}|^2\delta(\epsilon-\varepsilon_{k_\alpha}),\,\,\,
\alpha=L,R,
\end{equation}
and are assumed energy-independent, such that
$\Gamma_{\alpha}(\epsilon)\equiv\Gamma_{\alpha}$, $\alpha=L,R$. We count the
number of electrons that have been collected in the right lead, and
consistently with this choice, the counting field $\chi$ has been
introduced in the off-diagonal element of the memory kernel that
contains the rate $\Gamma_R$.

The expressions for the bath-assisted hopping rates
are derived in App.\ \ref{app:doubledot} and for real $z$ they
read
\begin{equation}
\Gamma_B^{(\pm)}(z)=T_c^2[g^{(+)}(z_\pm)+g^{(-)}(z_\mp )],
\end{equation}
where $z_\pm\equiv z\pm i\varepsilon+\Gamma_R/2$. These expression are valid to the lowest order in the tunnel coupling $T_c$. The
bath-correlation functions in Laplace space are
\begin{equation}
g^{(\pm)}(z)=\int_0^{\infty}dt e^{-W(\mp t)-zt}
\end{equation}
with\cite{Weiss2001}
\begin{equation}
W(t)=\int_0^{\infty}d\omega
\frac{J(\omega)}{\omega^2}\{[1-\cos(\omega
t)]\coth{(\beta\omega/2)}+i\sin(\omega t)\},
\end{equation}
and
\begin{equation}
J(\omega)\equiv\sum_j|g_j|^2\delta(\omega-\omega_j)
\end{equation}
being the spectral function of the heat bath. In this work we
consider Ohmic dissipation characterized by a coupling strength
$\alpha$ such that the spectral
function reads
\begin{equation}
J_\Omega(\omega)=2\alpha \omega e^{-\omega/\omega_c}.
\end{equation}
where $\omega_c$ is the frequency cut-off, assumed to be the highest energy scale of the system.

\begin{figure}
\begin{center}
\includegraphics[width=0.45\textwidth, trim = 0 0 0 0, clip]{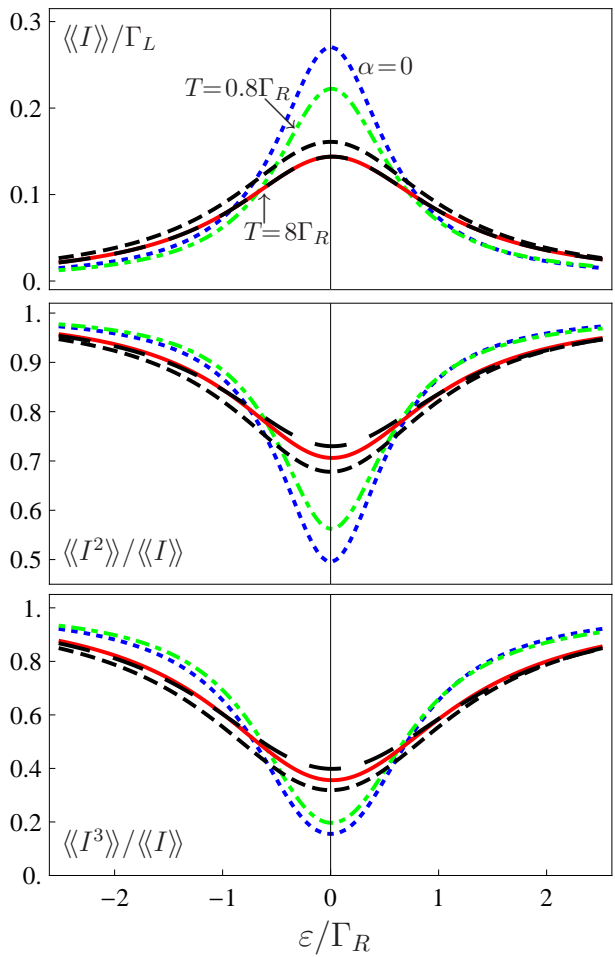}
\caption{(color online). Cumulants for weakly coupled heat bath.
The first three cumulants are shown as functions of the
dealignment $\varepsilon$. We show results for a low ($T=0.8\Gamma_R$)
and a high temperature ($T=8\Gamma_R$) as well as for the uncoupled
case ($\alpha=0$). For the high temperature case, we compare with
results obtained using a charge detector dephasing
model (short-dashed black line) and results without inclusion of memory
effects (long-dashed black line), see text. Parameters are
$\Gamma_L=0.1\Gamma_R$, $T_c=0.1\Gamma_R$, $\alpha=0.01$, and
$\omega_c=5\times10^4\Gamma_R$, and thus $\Gamma_d\simeq 0.8\Gamma_R$ according to Eq.\ (\ref{eq:dephaserate}).} \label{fig:DQDres1}
\end{center}
\end{figure}

In Fig.\ \ref{fig:DQDres1} we show results for weak couplings to
the heat bath, $\alpha\ll 1$. As the two quantum dot levels are
tuned into resonance ($\varepsilon=0$), the current reaches a
maximum with a width mainly determined by $\Gamma_R$. The
corresponding values for the second and third cumulant, normalized
with respect to the current, are suppressed below unity. The
suppression is stronger for the third cumulant. Away from
resonance, the mean current falls off, and the second and third
cumulants approach unity, corresponding to a Poisson process.
Without coupling to the heat bath, $\alpha=0$ (dotted blue line), the only broadening
mechanism is the escape of electrons through the right barrier at
rate $\Gamma_R$. This rate also defines the relevant energy scale to which we compare
the temperature of the heat bath. Away from resonance, the uncoupled case captures
well the results obtained at low temperatures, $T< \Gamma_R$. (dashed-dotted green line). At higher temperatures, $T\gg\Gamma_R$ (full red line), the peak in the current and the dips in the second and third cumulants are considerably
broadened due to the strong temperature induced dephasing.
Further results for the weak coupling limit are presented in Ref.~\onlinecite{Braggio2008}.

In order to understand the behavior at high temperatures (full red line), we
imagine replacing the heat bath by a charge detector which
measures the position of electrons on the DQD, thereby causing
dephasing.\cite{Gurvitz1997,Kiesslich2006,Braggio2009} The effects of the
charge detector can be described by a single dephasing rate
$\Gamma_d$, entering as an additional exponential decay of the
off-diagonal elements between the left and right quantum dot
states. As we show in App.\ \ref{app:doubledot}, this picture
follows from the high-temperature limit of the kernel in  Eq.\
(\ref{eq:kernel}), and the corresponding dephasing rate is
\begin{equation}
\Gamma_d=2\alpha \pi T.
\label{eq:dephaserate}
\end{equation}
The dynamics of the system effectively becomes Markovian at high temperatures $T\gg\Gamma_R$, where the characteristic memory time $\sim(\Gamma_R/2+\Gamma_d)^{-1}$ of the kernel is shorter than the timescale $\sim\Gamma_R^{-1}$ over which the populations of the DQD evolve.
In Fig.\ \ref{fig:DQDres1} we see that the counting statistics at
high temperatures (full red line) are well approximated  by the charge detector
model (short-dashed black line), which captures the broadening of the peak in the current
and the dips in the second and third cumulants. For high
temperatures (full red line), the large value of the dephasing rate indicates that
the system is strongly dephased. The charge detector model,
however, cannot account for the weak asymmetry between the phonon
emission ($\varepsilon>0$) and absorption ($\varepsilon<0$) sides at low temperatures (dashed-dotted green line).

In our description of the DQD system we have traced out the
electronic off-diagonal elements, the coherencies, together with
the electronic leads and the heat bath. Our derivation allows us to combine
strong coupling to the heat bath with broadening of the electronic levels due to the electrodes.
However, even without coupling to the heat bath, the kernel must be time-dependent in order to account for the coherent oscillations between the left
and right quantum dot states. These coherent effects are suppressed, when the dephasing is
strong, and in that limit we thus expect that a Markovian description would suffice. We check this assumption by
plotting in Fig.\ \ref{fig:DQDres1} only the Markovian parts of
the cumulants at high temperatures (long-dashed black line). Away from resonance, the
Markovian parts agree well for the second and third cumulants
showing that the system at high temperatures effectively is
Markovian. The mean current, as previously mentioned, is already a
Markovian quantity and it coincides with the Markovian
contribution as expected.\cite{Braggio2006} Closer to resonance,
some deviations for the second and third cumulants are seen as the system is not completely dephased.

\begin{figure}
\begin{center}
\includegraphics[width=0.45\textwidth, trim = 0 0 0 0, clip]{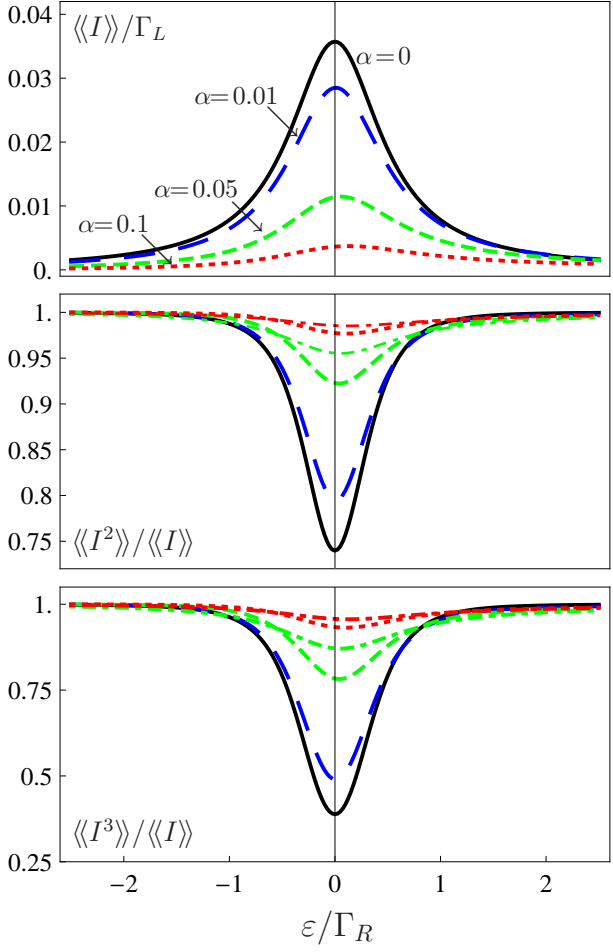}
\caption{(color online). Cumulants in the intermediate regime
between weak and strong coupling to the heat bath. The first three
cumulants are shown as functions of the dealignment $\varepsilon$.
The values of the coupling to the heat bath are $\alpha=0$, 0.01,
0.05, and 0.1. For large couplings ($\alpha=0.5,1$), we compare
with results obtained without inclusion of memory
effects (dot-dashed lines), see text. Parameters are
$\Gamma_L=\Gamma_R$, $T_c=0.1\Gamma_R$, $T=0.5\Gamma_R$, and
$\omega_c=5\times 10^4\Gamma_R$.}
\label{fig:DQDres2}
\end{center}
\end{figure}

In Fig.\ \ref{fig:DQDres2} we show results for larger values of
the coupling to the heat bath. Also in this case, the peak in the
current and the dips in the second and third cumulants are
suppressed as the coupling is increased and dephasing becomes
stronger. Contrary to the weak coupling regime, however, no
broadening of these features are observed. This is not consistent
with the charge detector model, which would predict an increased
width together with suppression of the height of the current peak
and the depth of the dips in the second and third cumulants.
Additionally, as the coupling $\alpha$ is increased, the
emission/absorption asymmetry becomes stronger as exchange of
energy quanta with the heat bath becomes increasingly important.
As noted before, neither the absence of broadening nor the
asymmetry of the peaks and dips can be accounted for by the
charge detector model. For large values of the coupling to the
heat bath, the DQD system completely dephases and the Markovian
contribution describes well the behavior of the counting
statistics, in particular away from resonance, where coherent
effects are less relevant. At even larger couplings, the heat bath
tends to localize electrons to one of the two quantum dots, and
the effective tunnel rate between the two quantum dots becomes
highly suppressed. In that case, the current through the system is
very low and the statistics is Poissonian.

\subsection{Discussion of non-Markovian systems}
\label{subsec:nonMarkcorr}

We close this section by pointing out various possible subtleties associated with
Markovian\cite{Spohn1979,Kohen1997} and, in particular,
non-Markovian\cite{Stenholm2001} GMEs. As we shall argue,
special attention should be paid to the
sometimes paradoxical nature of heuristically derived non-Markovian
GMEs (see e.\ g.\ Ref.~\onlinecite{Stenholm2001}) in order to ensure physically meaningful results.
We moreover discuss the interpretation of the ``mean memory time" for non-Markovian systems and we show that it in certain cases can turn negative. As an illustrative example, we consider unidirectional transport through a single electronic level. This generic model provides us with a unifying explanation of previous results obtained for several different systems and, in addition, it displays a few possible peculiarities of non-Markovian transport.

The Markovian master equation for the simple two-state model of unidirectional transport through a single
electronic level is determined by the rate matrix\cite{Blanter2000,Bagrets2003,Gustavsson2006,Flindt2009}
\begin{equation}\W(\chi)=
  \begin{pmatrix}
    -\Gamma_L & \Gamma_R e^{i\chi} \\
    \Gamma_L  & -\Gamma_R
  \end{pmatrix}
\end{equation}
with the counting field $\chi$ corresponding to tunneling across the
right barrier. An intuitive way to generalize the rate matrix to
the non-Markovian case would be simply to replace the rates $\Gamma_{L/R}$
by time-dependent rates, so that the rate matrix in Laplace
space instead reads
\begin{equation}
\label{eq:twostatekernel}
  \W(\chi,z)=
  \begin{pmatrix}
    -\Gamma_L(z) & \Gamma_R(z) e^{i\chi} \\
    \Gamma_L(z) & -\Gamma_R(z)
  \end{pmatrix}.
\end{equation}
The non-Markovian character could be caused by
external degrees of freedom  that have
been traced out, for example a
harmonic oscillator mode coupled to the occupation of the
electronic level, as studied in Sec.~\ref{subsec:vibration}.
Alternatively, it could be due to an energy-dependent tunneling density of states as in
Ref.~\onlinecite{Zedler2009}, or many-body-induced effects
as in the Fermi-edge-singularity problem in
transport.\cite{Matveev1992,Hapke-Wurst2000,Frahm2006,Maire2007,Ruth2008}

The microscopic origin of the memory, however, is not important in the following and its
effect on the current noise is qualitatively captured by the ``mean
memory times"
\begin{equation}
\tau_{L,R}\equiv-\left.\tfrac{d}{dz}\log\Gamma_{L,R}(z)\right|_{z=0}=\frac{\int_0^{\infty}t\Gamma_{L,R}(t)dt}{\int_0^{\infty}\Gamma_{L,R}(t)dt}.
\label{eq:memtime}
\end{equation}
If $\Gamma(t)\propto e^{-t/\tau}$ the mean memory time has a direct physical interpretation as the characteristic memory time $\tau$. However, the following statements are valid for any memory kernel as long as it has a finite mean memory time $\tau$. Using Eqs.\ (\ref{eq:nonMarkovcumulants},\ref{eq:nonMarkovcoeff}) we find the following expression for the
Fano factor [see also Eq.\ (13) of Ref.\ \onlinecite{Flindt2007}]
\begin{equation}
\label{eq:twostate}
  F=\frac{\llangle I^2\rrangle}{\llangle
  I\rrangle}=\frac{\Gamma_L^2+\Gamma_R^2}{(\Gamma_L+\Gamma_R)^2}-2\frac{\Gamma_L^2\Gamma_R\tau_R+\Gamma_R^2\Gamma_L\tau_L}{(\Gamma_L+\Gamma_R)^2}
\end{equation}
where $\Gamma_{L,R}\equiv\Gamma_{L,R}(z=0)>0$ is the Markovian limit
of the rates. The second term is a non-Markovian correction to the
well-known expression for the Fano factor of transport through a
single electronic level.\cite{Blanter2000} For certain parameter
values, however, this non-Markovian correction can make the Fano factor turn
negative --- a clearly unphysical result (the zero-frequency noise
must be positive, see e.g. Refs.~\onlinecite{Blanter2000,Flindt2004}). We discuss
this issue in further detail below.

Obviously, different baths producing the same mean
memory times give rise to the same electronic noise. Intuitively, one would also expect the mean
memory times to be positive.  Since the Markovian limit of the $\Gamma$'s must be
positive (being rates) the non-Markovian correction appears
negative; thus the general effect of memory on transport through a single level
is a decrease of the noise compared to the corresponding
Markovian limit. This statement is in line with a number of previous findings: It explains the anomalous suppression of the Fano factor below $1/2$ in
transport through a single electronic level coupled to a mechanical
resonator reported in Ref.~\onlinecite{Haupt2006}. Additionally, the
non-Markovian correction due to strong spectral features in the
Fermi edge singularity
problem\cite{Matveev1992,Hapke-Wurst2000,Frahm2006,Ruth2008} is
responsible for the observed discrepancy\cite{Maire2007} between the
measured Fano factor and the expected result based on the Markovian
part only. A more
detailed account of this problem will be presented
elsewhere.\cite{Roszakinprep} Finally, the suppression of the Fano
factor compared to the Markovian approximation is also confirmed by
the study of an exactly solvable case\cite{Zedler2009} in the regime
where the non-Markovian GME provides a good approximation to the
exact dynamics.

Although the above expression (\ref{eq:twostate}) provides a
unifying explanation of these three examples, it obviously cannot be
correct in general as mentioned above. For weak non-Markovian behavior, where the $\tau$'s are small, the Fano factor stays positive, and the non-Markovian corrections lead to a reduction of noise. In general, however, there is no guaranty that the
Fano factor in Eq.\ (\ref{eq:twostate}) is always non-negative. This can be traced back to the heuristic
inclusion of the non-Markovian kernel in Eq.\
(\ref{eq:twostatekernel}). While the non-Markovian kernel for
unidirectional transport through a single level in general may be
written in the form \eqref{eq:twostatekernel} {\em without} counting
fields, the inclusion of the counting field must be carried out
carefully, for example by using well-controlled
systematic derivation procedures, such as those based on
perturbation theories.\cite{Braggio2006}
Non-Markovian GMEs may, however, still lead to unphysical results, when employed outside their regime of validity, as recently discussed by Zedler and co-workers.\cite{Zedler2009} In
the Markovian case, the heuristic addition of the counting field in $\W$ usually leads to correct results for the counting statistics, although exceptions do exist.\cite{Prachar2009} In the non-Markovian double dot system studied in Sec.\ \ref{subsec:DQD}, the counting field enters the Markovian ($z$-independent) part of kernel in Eq.\ (\ref{eq:kernel}), and the inclusion of the counting field does not lead to any of the issues discussed above.

Finally, we discuss another subtlety associated with strongly non-Markovian
systems. Under certain circumstances the mean memory time $\tau$,
defined in Eq.\ (\ref{eq:memtime}) may in fact become negative. This
happens for example for the dissipative double quantum dot studied in Sec.~\ref{subsec:DQD}.
For a sufficiently small dissipation rate, we find for the bath-assisted rates $\partial_z\Gamma_B^{(\pm)}(z=0)>0$, resulting in negative mean memory times. Formally, there is no problem associated with this phenomenon (the GME still describes a
positivity-preserving evolution), but the physical interpretation of the non-Markovian corrections is less clear due to this counterintuitive behavior.  The problem is purely
interpretation-related and concerns the issue of a proper Markovian
limit.

In cases with large memory effects, the formal Markovian limit, corresponding to $\W(z\rightarrow 0)$,  does
not give a reasonable description of the system {\em dynamics}, although it yields correct
{\em stationary} quantities, like the mean current.
Since the noise (and also higher-order cumulants) is a time
integral of a transient quantity, namely a current-current
correlation function, the formal Markovian limit of the noise in these cases is a somewhat
unphysical quantity. The problems with the interpretation of a Markovian
limit also influence the interpretation of the non-Markovian
corrections (via, e.\ g.\ negative memory times). Bluntly, a
physically meaningful result is
arbitrarily split into two additive parts, Markovian and non-Markovian, that each do not necessarily have a reasonable physical interpretation. The full result, however, is correct and
physically plausible. These effects are well
illustrated and can be understood by studying exactly solvable cases such
as the one in Ref.~\onlinecite{Zedler2009}.

In this section, we have only briefly touched upon various open questions and subtleties associated with interpretations of non-Markovian dynamics. However, the exact method developed in this paper paves the way for future systematic studies of memory effects in connection with electronic noise and counting statistics.

\section{Conclusions}
\label{sec:conclusions}

We have presented a detailed derivation of a recursive scheme for evaluating high-order cumulants of transport through Coulomb-blockade nanostructures with many states and non-Markovian dynamics. In order to illustrate the use of our method for Markovian systems we considered the counting statistics of transport through a two-level quantum dot and a vibrating molecule. In both cases, we have shown how the behavior of high-order cumulants is determined by dominating singularities of the cumulant generating functions. Oscillations of the high-order cumulants as function of the cumulant order can be used to locate the positions of singularities as we have demonstrated. We have also calculated the distribution of measurable currents, the so-called large deviation function, and shown how the tails of the distributions reflect the high-order cumulants. In order to illustrate the use of our method for a non-Markovian system, we considered transport through a dissipative double quantum dot. For this system, we have studied how bath-induced dephasing affects the first three cumulants and found that effects of the heat bath cannot be accounted for by an effective detector model, when the coupling becomes strong. Finally, we have discussed the nature and significance of non-Markovian dynamics in relation to counting statistics.

The research presented in this work points to several interesting directions to follow. While we have focused on the zero-frequency current cumulants of non-Markovian processes, it would be interesting to see, if the methods presented here could be extended to finite frequencies, as it was recently done for Markovian processes.\cite{Emary2007} It has now been firmly established that high-order cumulants of the counting statistics generally grow factorially with the cumulant order and oscillate as functions of basically any system parameters as well as of the cumulant order. It would be interesting to study in further detail how microscopic details of a system are reflected, for example, in the frequency of these oscillations. Such a study would shed new light on the information contained in high-order cumulants.
Finally, we believe that the methods presented here will pave the way for future systematic studies of counting statistics in connection with non-Markovian dynamics.

\acknowledgements

We thank R.~Aguado, T.~Brandes, C.\ Emary, D.\ Kambly, S.~Kohler, D.~Marcos, K.~Neto\v{c}n\'{y},
M.~Sassetti, P.~Talkner, J.\ Zamastil, and P.\ Zedler  for fruitful discussions and
suggestions. We thank the group of R.\ J.\ Haug for enlightening discussions about
experimental aspects of counting statistics. The work was supported by the Villum Kann Rasmussen
Foundation, INFM--CNR Seed Project, European Science Foundation
(`Arrays of Quantum Dots and Josephson Junctions'), Czech Science
Foundation (grant 202/07/J051), and FiDiPro of the Finnish Academy.
The work of T.~N.\ is a part of the research plan MSM 0021620834
financed by the Ministry of Education of the Czech Republic.

\appendix

\section{QR decomposition}
\label{app:QR}

In this appendix we present technical details of one possible method
for evaluating Eqs.~\eqref{eq:hom_eq} and \eqref{eq:Req} in
Sec.~\ref{subsec:evaluation}. The method we use is a standard technique in
numerical linear algebra known as QR decomposition. Routines
performing QR decompositions are a part of most common linear algebra
packages such as {\tt LAPACK}. Below we provide a piece
of code implemented in {\tt MATLAB} and subsequently explain in detail each step of the code,
such that it can be reproduced in other programming
languages ({\tt MATLAB} itself uses an implementation of the QR
decomposition from the underlying {\tt LAPACK} library). The code
is written in a general way but there are steps which are
specific to the particular model considered here --- those program lines are explicitly
denoted. As a model system we use a double dot with 5 retained
elements of the density matrix $\hat{\rho}$ and corresponding vector representation
\begin{equation}
|\rho\rrangle=[\rho_{00},\rho_{LL},\rho_{RR},\rho_{LR},\rho_{RL}]
\end{equation}.
The trace of the density matrix can then be written
\begin{equation}
\mathrm{Tr}\{\hat{\rho}\}=\llangle\tilde{0}|\rho\rrangle=\rho_{00}+\rho_{LL}+\rho_{RR}
\end{equation}
with
\begin{equation}
\llangle\tilde{0}|=[1,1,1,0,0]^T.
\end{equation}

The excerpt of the code for the evaluation of the stationary state
Eq.~\eqref{eq:hom_eq} and the pseudo-inverse (modification of
Eq.~\eqref{eq:Req})
\begin{equation}\label{eq:pseudoinverse}
    \mathcal{W R}=\mathcal{Q}
\end{equation}
reads:

\begin{verbatim}
% Size of the Liouville space
N = length(W(:,1));

% Left zero eigenvector (MODEL DEPENDENT)
%  Here for a double dot
trace=[1,1,1,0,0];

% QR decomposition with sorting the diagonal
%  elements of 'r' in descending order
%  (built-in routine from LAPACK)
[q,r,e] = qr(W);

% Consistency check - when the matrix 'W' is
%  singular the last row of the matrix 'r'
%  should be zero
tol = 1e-10;    % Setting the tolerance
if max(r(end,:))>tol
     warning('Last row of r is non-zero')
     display(r(end,:))
end

% Stationary state
stat = e*[r(1:N-1,1:N-1) \ r(1:N-1,end); -1];

% Normalization of the stationary state
stat = stat/(trace*stat);

% Projectors
P = kron(stat,trace); Q = eye(N) - P;

% Solution of the pseudo-inverse equation
temp = q \ Q;

% Consistency check - if the matrix 'Q' is
%  a projector onto the regular space of 'W',
%  the solvability condition requires the
%  last column of the matrix 'temp' to be zero
if max(temp(end,:))>tol
    warning('Last row is non-zero')
    display(max(temp(end,:)))
end

% Finding a particular solution of the equation
X = e*[r(1:N-1,1:N-1)\temp(1:N-1,:);zeros(1,N)];

% Final fixing of the pseudo-inverse by
%  multiplication by the projector 'Q'
R = Q*X;
\end{verbatim}

The code can be understood by first analyzing the structure of the output from the QR decomposition routine. The QR routine takes as input a matrix $\mathbf{W}$
of dimension $N\times N$. The output consists of three matrices of same dimension
$\mathbf{q},\,\mathbf{e},\,\mathbf{r}$, such that $\mathbf{q}\cdot
\mathbf{r} = \mathbf{W}\cdot \mathbf{e}$. Here, $\mathbf{q}$ is a
unitary (and thus regular) matrix, $\mathbf{e}$ a permutation
matrix (thus also regular), and $\mathbf{r}$ an upper
triangular matrix with decreasing diagonal elements. The column
permutation matrix $\mathbf{e}$ is chosen such that
$\mathrm{abs}(\mathrm{diag}(\mathbf{r}))$ is decreasing. For a
singular matrix $\mathbf{W}$ representing $\mathcal{W}$, this implies that the last diagonal entry of $\mathbf{r}$ is
zero and, therefore, the last row of $\mathbf{r}$ is zero.
More explicitly, we have
\begin{subequations}
\begin{align}
   \mathbf{r} & = \begin{pmatrix}
        r_{1,1} & r_{1,2} & r_{1,3} & r_{1,4} & \dots & r_{1,N} \\
        0 & r_{2,2} & r_{2,3} & r_{2,4} & \dots & r_{2,N} \\
        0 & 0 & r_{33} & r_{3,4} & \dots & r_{3,N} \\
        \vdots & \vdots & \vdots & \ddots & \ddots & \vdots \\
        0 & 0  &  \dots & 0 & r_{N-1,N-1} & r_{N-1,N} \\
        0 & 0 & 0 & 0  & 0 & 0 \\
        \end{pmatrix} \\
    & =\begin{pmatrix} [\mathbf{\tilde{r}}]_{N-1\times N-1} & [\mathbf{r'}]_{N-1\times 1} \\ [0]_{1\times N-1} & 0 \end{pmatrix}\label{eq:rmatrix}
\end{align}
\end{subequations}
with $\mathbf{\tilde{r}}$ being an upper triangular matrix with
non-zero diagonal and dimension $(N-1)\times (N-1)$ and $\mathbf{r'}$ a
column vector of length $N-1$. The QR decomposition of $\mathcal{W}$ implies
for the solution $|0\rrangle$ to the matrix implementation of
Eq.~\eqref{eq:hom_eq} that
\begin{equation}
    \mathbf{W}\cdot\mathbf{0} = \left(\mathbf{q}\cdot \mathbf{r}\cdot \mathbf{e}^{-1}\right)\cdot\mathbf{0}=0 \ \Rightarrow\
    \mathbf{r}\cdot\left(\mathbf{e}^{-1}\cdot\mathbf{0}\right)=0.
\end{equation}
Here, $\mathbf{0}$ is a vector representation of $|0\rrangle$, named {\tt stat} in the code.

The block structure of the matrix $\mathbf{r}$ depicted in Eq.\ \eqref{eq:rmatrix}
shows that $(\mathbf{e}^{-1}\cdot\mathbf{0}) = c[\mathbf{\tilde{r}}^{-1}\cdot \mathbf{r'},-1]^T$ for any number $c$ is a solution to the matrix equation above. We then find $\mathbf{0} = c\mathbf{e}\cdot[\mathbf{\tilde{r}}^{-1}\cdot \mathbf{r'},-1]^T$ with  $c=1/(\mathbf{\tilde{0}}\cdot\mathbf{e}\cdot[\mathbf{\tilde{r}}^{-1}\cdot \mathbf{r'},-1]^T)$, ensuring the proper normalization $\mathbf{\tilde{0}}\cdot\mathbf{0}=1$. We note that the
only model dependent part of the code is the definition of
the left zero eigenvector $\mathbf{\tilde{0}}$ named {\tt trace} in the code.

Next, we determine the pseudo-inverse $\mathbf{R}$. To this end, we form the projectors $\mathbf{P}=\mathbf{0}\otimes\mathbf{\tilde{0}}$ and $\mathbf{Q}=\mathbf{1}-\mathbf{P}$. For the Kronecker tensor product, the code uses the built-in function {\tt kron}(,).
Equation \eqref{eq:pseudoinverse} can now be expressed in matrix form as
\begin{equation*}
   \mathbf{W}\cdot\mathbf{R} = \left(\mathbf{q}\cdot \mathbf{r}\cdot \mathbf{e}^{-1}\right)\cdot\mathbf{R}=\mathbf{Q} \ \Rightarrow\
    \mathbf{r}\cdot\left(\mathbf{e}^{-1}\cdot\mathbf{R}\right)=\mathbf{q}^{-1}\cdot\mathbf{Q}
\end{equation*}
Since the right hand side $\mathcal{Q}$ of the original equation
\eqref{eq:pseudoinverse} lies in the range of $\mathcal{W}$ the
resulting matrix equation for
$\left(\mathbf{e}^{-1}\cdot\mathbf{R}\right)$ above must have a
solution. This requires that the last row of
$\mathbf{q}^{-1}\cdot\mathbf{Q}$ is zero as follows
again from the block structure of $\mathbf{r}$ shown in Eq.\ \eqref{eq:rmatrix}
(this condition is explicitly checked for in the code). A
particular solution of the equation is then
$\begin{pmatrix}\mathbf{\tilde{r}}^{-1}\cdot[\mathbf{A}]_{N-1\times N}\\
[0]_{1\times N}\end{pmatrix}$ with the rectangular matrix
$\mathbf{A}$ being a restriction of the product
$\mathbf{q}^{-1}\cdot\mathbf{Q}$ to the $(N-1)$ first rows. The pseudo-inverse is now fixed
by multiplying this particular solution by the permutation matrix
$\mathbf{e}$ and finally by the projector $\mathbf{Q}$ as follows
from the discussion below Eq.~\eqref{eq:RRx}.

We have used the code  on
a standard PC for all of the examples shown in this paper and it
is efficient both in terms of memory and CPU time for $N\times N$ matrices with $N$ being up to several thousands. For larger matrices, the direct evaluation of the pseudo-inverse becomes
prohibited both in terms of memory as well as CPU time and
one should use other methods such as, e.g., the iterative Arnoldi
scheme described in detail in App.\ A of
Ref.~\onlinecite{Flindt2004}.

\section{Vibrating molecule}
\label{app:vibmol}

In this appendix we describe the model of a vibrating molecule considered in Sec.\ \ref{subsec:vibration} and derive the corresponding Markovian GME. Here we follow to
a large extent the description of the model given in Refs.\
\onlinecite{McCarthy2003}, \onlinecite{Flindt2005}. The molecule is operated in
the Coulomb blockade regime, where only two charge states ($m=0$ or
$m=1$ additional electrons on the molecule) participate in the
transport. We consider spinless electrons with charge $-e$, although
it would be easy to include the spin degree of freedom. The
Hamiltonian of the molecule is
\begin{equation}
\hatH_S =
\frac{\hat{p}^2}{2m_0}+\frac{1}{2}m_0\omega_0^2\hat{x}^2+(\varepsilon-eE\hat{x})\hat{d}^{\dagger}\hat{d},
\end{equation}
where $E$ is the electric field at the position of the molecule with
mass $m_0$ and natural oscillator frequency $\omega_0$. The electric
field is determined by the bias across the molecule and
bias-independent contributions, e.g., image-charge effects. The
charging energy difference between 0 and 1 additional electron on
the molecule is denoted as $\varepsilon$. The molecule is
tunnel-coupled to left and right electrodes consisting of
non-interacting fermions
\begin{equation}
\hatH_{\alpha}=\sum_{k_\alpha}\varepsilon_{k_\alpha}\hat{c}^{\dagger}_{k_\alpha}\hat{c}_{k_\alpha},\,\,
\alpha=L,R, \end{equation} kept at chemical potentials
$\mu_{\alpha}$, $\alpha=L,R$, and temperature $T$. Tunneling processes are accounted for
by a standard tunnel-Hamiltonian
\begin{equation}
\hatH_T=\sum_{k_{\alpha},\alpha=L,R}\left(t_{k_{\alpha}}\hat{c}^{\dagger}_{k_\alpha}\hat{d}+\mathrm{h.c.}\right).
\end{equation}
For simplicity, we neglect any position dependence of the tunneling
amplitudes $t_{k_\alpha}$, but it would be straightforward to
include.\cite{McCarthy2003} Finally, damping of the mechanical
oscillations are described by coupling to a bath of oscillators,
such that the full Hamiltonian reads
\begin{equation}
\hatH = \hatH_S+\hatH_T+\hatH_L+\hatH_R+\hat{x}\hat{V}_B+\hatH_B,
\end{equation}
where
\begin{equation}
\hat{V}_B=\sum_j\frac{g_j}{2}(\hat{a}^{\dagger}_j+\hat{a}_j),
\end{equation}
and
\begin{equation}
\hat{H}_B=\sum_j\hbar\omega_j\hat{a}^{\dagger}_j\hat{a}_j.
\end{equation}
The coupling to the $j$'th oscillator, with frequency $\omega_j$ and
corresponding creation and annihilation operators
$\hat{a}^{\dagger}_j$ and $\hat{a}_j$, respectively, is denoted
$g_j$.

We treat both the coupling to the electronic leads and the heat bath
in the weak coupling approximation and it thus suffices to consider
the time evolution of the diagonal matrix elements of the reduced
density matrix of the charge and oscillator states of the molecule.
These diagonal elements correspond to the energy eigenstates of the
isolated molecule described by $\hatH_S$. The eigenstates with $m=0$
additional electrons on the molecule are
\begin{equation}
|m=0,l\rangle = |\mathrm{empty}\rangle\otimes|l\rangle
\label{eq:eigenstatesempty}
\end{equation}
with corresponding eigenenergies
\begin{equation}
E_{0l}=\hbar\omega_0\left(l+\frac{1}{2}\right).
\end{equation}
Here, $|\mathrm{empty}\rangle$ denotes the empty charge state, while
$|l\rangle=(\hat{a}^{\dagger})^l|0\rangle/\sqrt{l!}$ is the $l$'th oscillator
state centered at $x=0$. The operator $\hat{a}^{(\dagger)}$ lowers
(raises) the oscillator number by 1 and $|0\rangle$ is the
oscillator ground state. With $m=1$ additional electron on the
molecule the equilibrium position of the oscillator is shifted by
the distance $d=eE/m_0\omega_0^2$. The eigenstates for the occupied
molecule are thus
\begin{equation}
|m=1,l\rangle = |\mathrm{occupied}\rangle\otimes
e^{\gamma(\hat{a}^{\dagger}-\hat{a})}|l\rangle,
\label{eq:eigenstatescharged}
\end{equation}
where we have introduced the dimensionless electron-phonon coupling
\begin{equation}
\gamma=\frac{eEx_0}{\hbar\omega_0}
\end{equation}
with $x_0=\sqrt{\hbar/2m_0\omega_0}$. The corresponding
eigenenergies are
\begin{equation}
E_{1l}=\varepsilon+\hbar\omega_0\left(l+\frac{1}{2}\right)-\gamma^2\hbar\omega_0.
\end{equation}

In the following we denote the diagonal elements of the reduced
density matrix by $\rho_{m,l}(n,t)$, where $n$ is the number of
electrons collected in the right electrode during the time span
$[0,t]$. Bath-mediated transitions between different vibrational
states are given by the thermal rates\cite{McCarthy2003}
\begin{equation}
W_{l+1\leftarrow l}=W_{l\leftarrow
l+1}e^{-\beta\hbar\omega_0}=K\frac{\hbar(l+1)}{e^{\beta\hbar\omega_0}-1},
\end{equation}
where $K$ characterizes the vibrational damping rate and
$\beta=1/k_BT$ is the inverse temperature. The charge transfer
rates, obtained using Fermi's Golden rule, are
\begin{equation}
\begin{split}
\Gamma_{1,l'\leftarrow 0,l}^{(s)} &=
\Gamma_{(s)}|F_{l'l}|^2f(E^{(s)}_{ll'}),
\\
\Gamma_{0,l\leftarrow 1,l'}^{(s)} &=
\Gamma_{(s)}|F_{ll'}|^2[1-f(E^{(s)}_{ll'})],
\end{split}
\end{equation}
where $f$ is the Fermi function, $\Gamma_{(+1/-1)}=\Gamma_{L/R}$ are
the bare tunneling rates, which are assumed to be energy
independent, i.\ e.\
\begin{equation}
\Gamma_{\alpha}=\Gamma_{\alpha}(\epsilon)=2\pi\sum_{k_\alpha}|t_{k_\alpha}|^2\delta(\epsilon-\varepsilon_{k_\alpha}),\,\,\alpha=L,R.
\end{equation}
Moreover, we have defined
\begin{equation}
\begin{split}
E^{(s)}_{ll'}=&E_{1l'}-E_{0l}+\frac{seV}{2}\\
=&\varepsilon+\frac{seV}{2}+\hbar\omega_0(l'-l-\gamma^2)
\end{split}
\end{equation}
with $V$ being the symmetrically applied bias, such that
$\mu_L=eV/2$ and $\mu_R=-eV/2$,  and the index $s$ indicating
whether an electron tunneled from/to the left ($s=-1$) or right
($s=+1$) lead. Finally, the matrix elements
\begin{equation}
F_{ll'}=\langle l|e^{\gamma(\hat{a}^{\dagger}-\hat{a})}|l'\rangle
\end{equation}
are the Franck-Condon overlaps between harmonic oscillator states
that have been shifted spatially with respect to each other due to
different charge occupations, cf.\ Eqs.\ (\ref{eq:eigenstatesempty})
and (\ref{eq:eigenstatescharged}). In the following, we assume that
the bias dependence of the electron-phonon coupling takes the form
\begin{equation}
\gamma = c_1+\frac{eV}{\hbar\omega_0}c_2
\end{equation}
with $c_1$ and $c_2$ being constants.\cite{McCarthy2003}

Having identified all relevant transition rates the
Markovian master equation for the diagonal elements of the reduced
density matrix $\rho_{m,l}(n,t)$ reads
\begin{widetext}
\begin{equation}
\begin{split}
\frac{d}{dt}\rho_{m,l}(n,t)=&-\left[\sum_{l'=l\pm 1}W_{l'\leftarrow
l}\,\,\,+\sum^{\infty}_{\substack{l'=0\\
s=\pm 1,m'=0,1}}\Gamma_{m',l'\leftarrow
m,l}^{(s)}\right]\rho_{m,l}(n,t)+\sum_{l'=l\pm 1}W_{l\leftarrow l'}\rho_{m,l'}(n,t)\\
&+\sum_{l'=0}^{\infty}\left[\Gamma_{m,l\leftarrow
1-m,l'}^{(-1)}\rho_{1-m,l'}(n,t)+\Gamma_{m,l\leftarrow
1-m,l'}^{(+1)}\left\{(1-m)\rho_{1,l'}(n-1,t)+m\rho_{0,l'}(n+1,t)\right\}\right].
\end{split}
\label{eq:modelC601}
\end{equation}
We introduce the counting field via the transformation
$\rho_{m,l}(\chi,t)=\sum_n\rho_{m,l}(n,t)e^{in\chi}$. The
corresponding master equation, obtained from Eq.\
(\ref{eq:modelC601}), reads
\begin{equation}
\begin{split}
\frac{d}{dt}\rho_{m,l}(\chi,t)=&-\left[\sum_{l'=l\pm
1}W_{l'\leftarrow
l}\,\,\,+\sum^{\infty}_{\substack{l'=0\\
s=\pm 1,m'=0,1}}\Gamma_{m',l'\leftarrow
m,l}^{(s)}\right]\rho_{m,l}(\chi,t)+\sum_{l'=l\pm 1}W_{l\leftarrow l'}\rho_{m,l'}(\chi,t)\\
&+\sum_{l'=0}^{\infty}\left[\Gamma_{m,l\leftarrow
1-m,l'}^{(-1)}\rho_{1-m,l'}(\chi,t)+\Gamma_{m,l\leftarrow
1-m,l'}^{(+1)}\left\{e^{i\chi}(1-m)\rho_{1,l'}(\chi,t)+e^{-i\chi}m\rho_{0,l'}(\chi,t)\right\}\right].
\end{split}
\label{eq:modelC602}
\end{equation}
The elements $\rho_{m,l}(\chi,t)$ are collected in the vector
$\hatrho(\chi,t)$, whose equation of motion reads
\begin{equation}
\frac{d}{dt}\hatrho(\chi,t)=\W(\chi)\hatrho(\chi,t).
\end{equation}
The matrix elements of $\W(\chi)$ are identified from Eq.\
(\ref{eq:modelC602}).

\section{Double dot system}
\label{app:doubledot}

In this appendix we derive the expression for the memory kernel
given in Eq.\ (\ref{eq:kernel}). We take as our starting point the full
Hamiltonian in Eq.\ (\ref{eq:fullHamiltonian}). Following Gurvitz
and Prager\cite{Gurvitz1996} we project out the electronic leads in
order to obtain an equation of motion for the reduced density matrix
$\hatsigma=(\hatsigma_{00},\hatsigma_{LL},\hatsigma_{RR},\hatsigma_{LR},\hatsigma_{RL})^T$
of the three electronic states $|0\rangle,|L\rangle, |R\rangle$, and
the bath of oscillators. The off-diagonal elements
$\hatsigma_{0\alpha}$ and $\hatsigma_{\alpha0}$, $\alpha=L,R$,
between states with different charge occupation numbers are
decoupled from the rest and can therefore be disregarded. Following
the procedure described in Ref.\ \onlinecite{Gurvitz1996} we find
\begin{equation}
\frac{d}{dt}\hatsigma(t)=
\begin{pmatrix}
  -\Gamma_L & 0     & \Gamma_R  & 0                       & 0 \\
  \Gamma_L  & 0     & 0         & iT_c                   & -iT_c \\
  0         & 0     & -\Gamma_R & -iT_c                    & iT_c \\
  0         & iT_c & -iT_c      & -i\varepsilon-\Gamma_R/2 & 0 \\
  0         & -iT_c  & iT_c     & 0                       &
  i\varepsilon-\Gamma_R/2
\end{pmatrix}
\hatsigma(t)\\
 -i\begin{pmatrix}
  [\hatH_B,\hatsigma_{00}(t)]  \\
  [\hatH_B,\hatsigma_{LL}(t)]  \\
  [\hatH_B,\hatsigma_{RR}(t)]  \\
  [\hatH_B,\hatsigma_{LR}(t)]  \\
  [\hatH_B,\hatsigma_{RL}(t)]
  \end{pmatrix}
-i
\begin{pmatrix}
    0  \\
  [\hat{V}_B,\hatsigma_{LL}(t)]  \\
  -[\hat{V}_B,\hatsigma_{RR}(t)]  \\
  \{\hat{V}_B,\hatsigma_{LR}(t)\}  \\
  -\{\hat{V}_B,\hatsigma_{RL}(t)\}
\end{pmatrix},
\label{eq:gurvitz}
\end{equation}
where curly brackets denote anti-commutators
$\{\hat{A},\hat{B}\}\equiv\hat{A}\hat{B}+\hat{B}\hat{A}$ and we have
taken $\hbar=1$. The equation is valid when a large bias is driving
electrons through the double dot from the left lead to the right
lead with energy-independent rates\cite{Gurvitz1996}
\begin{equation}
\Gamma_{\alpha}=2\pi\sum_{k}|t_{k_{\alpha}}|^2\delta(\epsilon-\varepsilon_{k_\alpha}),\,\, \alpha=L,R.
\end{equation}
At this stage, the expression is
valid to all orders in the tunnel coupling $T_c$. Due to the
large-bias assumption, the energy $\varepsilon_0$ of the `empty'
state $|0\rangle$  drops out of the problem.

We now define the electronic occupation probabilities
$\rho_{i}\equiv\mathrm{Tr}_B\{\hatsigma_{ii}\}$, $i=0,L,R$, where
$\mathrm{Tr}_B$ is a trace over the bosonic degrees of freedom. For
these probabilities we readily find
\begin{equation}
\begin{split}
\frac{d}{dt}\rho_{0}(t)&=-\Gamma_L\rho_{0}(t)+\Gamma_R\rho_{R}(t),\\
\frac{d}{dt}\rho_{L}(t)&=\Gamma_L\rho_{0}(t)-2T_c\mathrm{Im}\left[\mathrm{Tr}_B\{\hatsigma_{LR}(t)\}\right],\\
\frac{d}{dt}\rho_{R}(t)&=-\Gamma_R\rho_{R}(t)+2T_c\mathrm{Im}\left[\mathrm{Tr}_B\{\hatsigma_{LR}(t)\}\right].
\end{split}
\label{eq:eomP}
\end{equation}
We proceed by considering the equation of motion for
$\hatsigma_{LR}$ obtained from Eq.\ (\ref{eq:gurvitz})
\begin{equation}
\label{eq:cohdiffeq}
\frac{d}{dt}\hatsigma_{LR}(t)=-(i\varepsilon+\Gamma_R/2)\hatsigma_{LR}(t)-i[\hatH_B^{(+)}\hatsigma_{LR}(t)
-\hatsigma_{LR}(t)\hatH_B^{(-)}]+iT_c[\hatsigma_{LL}(t)-\hatsigma_{RR}(t)],
\end{equation}
having defined $\hatH_B^{(\pm)}\equiv \hatH_B\pm \hat{V}_B$. Its
solution formally reads
\begin{equation}
\hatsigma_{LR}(t)=iT_c\int_0^tdt'e^{-(i\varepsilon+\Gamma_R/2)(t-t')}e^{-i\hatH_B^{(+)}(t-t')}
\left[\hatsigma_{LL}(t')-\hatsigma_{RR}(t')\right]e^{i\hatH_B^{(-)}(t-t')}+e^{-(i\varepsilon+\Gamma_R/2)t}
e^{-i\hatH_B^{(+)}t}\hatsigma_{LR}(0)e^{i\hatH_B^{(-)}t}.
\label{eq:sigmaLR}
\end{equation}
The first term enters the memory kernel below, while the second term
enters the inhomogeneity. In order to obtain a closed system of
equations for the three probabilities in Eq.\ (\ref{eq:eomP}), we
assume that the bath of oscillators between each tunneling event
reaches a local equilibrium corresponding to the given charge state.
This corresponds to the decoupling
\begin{equation}
\begin{split}
\label{eq:statedepmodel}
\hatsigma_{LL}(t)&\simeq \rho_{L}(t)\otimes\hatsigma^{(+)}(\beta),\\
\hatsigma_{RR}(t)&\simeq \rho_{R}(t)\otimes\hatsigma^{(-)}(\beta)
\end{split}
\end{equation}
in Eq.\ (\ref{eq:sigmaLR}), where
\begin{equation}
\hatsigma^{(\pm)}(\beta)\equiv
e^{-\beta H_B^{(\pm)}}/\mathrm{Tr}_B\{e^{-\beta H_B^{(\pm)}}\}
\end{equation}
and $\beta=1/k_BT$ is the inverse temperature.
The approximation is valid to lowest order in  $T_c^2$. Note that
no Markov approximation is made in this step. We then find
\begin{equation}
\begin{split}
\frac{d}{dt}\rho_{0}(t)&=-\Gamma_L\rho_{0}(t)+\Gamma_R\rho_{R}(t),\\
\frac{d}{dt}\rho_{L}(t)&=\Gamma_L\rho_{0}(t)
-\int_{0}^tdt' \left[\Gamma_B^{(+)}(t-t')\rho_L(t')-\Gamma_B^{(-)}(t-t')\rho_R(t')\right]-\gamma(t),\\
\frac{d}{dt}\rho_{R}(t)&=-\Gamma_R\rho_{R}(t)
+\int_{0}^tdt'\left[\Gamma_B^{(+)}(t-t')\rho_L(t')-\Gamma_B^{(-)}(t-t')\rho_R(t')\right]+\gamma(t),
\end{split}
\label{eq:nonMarkovEOM}
\end{equation}
where the inhomogeneity is of the form
$\hatgamma=(0,-\gamma,\gamma)^T$ and the bath-assisted hopping
rates are defined and evaluated below. An explicit expression for
the inhomogeneity will not be given in this work as we are only
considering the long-time limit for which the inhomogeneity is
irrelevant. By switching to Laplace space, the memory kernel given
in Eq.\ (\ref{eq:kernel}) is identified from Eq.\
(\ref{eq:nonMarkovEOM}) after the counting field has been incorporated via the substitution $\Gamma_R\rightarrow \Gamma_Re^{i\chi}$ in the first line of Eq.\
(\ref{eq:nonMarkovEOM}). In this example, the counting field enters the Markovian part of the kernel, and we do not encounter any of the problems described in Sec.\ \ref{subsec:nonMarkcorr}.
\end{widetext}
In the equations above we have defined the bath-assisted hopping
rates
\begin{equation}
\Gamma_B^{(\pm)}(t)\equiv2T_c^2\mathrm{Re}\left[e^{-(i\varepsilon+\Gamma_R/2)t}g^{(\pm)}(t)\right]
\label{eq:bathassistedrates}
\end{equation}
in terms of the bath correlation functions
\begin{equation}
\label{eq:gbath}
\begin{split}
g^{(\pm)}(t)&\equiv\mathrm{Tr}_B\{ e^{-i \hatH_B^{(+)}
t}\hatsigma^{(\pm)}(\beta) e^{i \hatH_B^{(-)}t}\}\\
&\equiv\left\langle e^{i \hatH_B^{(-)}t}e^{-i \hatH_B^{(+)}
t}\right\rangle_{(\pm)}.
\end{split}
\end{equation}
These bath correlation functions can be evaluated using standard
many-particle techniques.\cite{Mahan1990} First we introduce a
polaron transformation of the form
\begin{equation}
\label{eq:PolaronS}
\hat{S}^{(\pm)}=e^{\pm
i\hat{A}},\,\,\,\hat{A}=\sum_j\frac{ig_j}{2\omega_j}(\hat{a}_j-\hat{a}^{\dagger}_j),
\end{equation}
which removes $\hat{V}_B$ from the bath correlation functions, since
\begin{equation}
\label{eq:PolaronProp}
\hat{S}^{(\pm)}\hatH_{B}^{(\pm)}[\hat{S}^{(\pm)}]^\dagger=\hatH_{B}-\frac{g_j^2}{4\omega_j}.
\end{equation}
By insertion of the identity $1^{(\pm)}=\hat{S}^{(\pm)}[\hat{S}^{(\pm)}]^\dagger\equiv 1
$ in Eq.\ (\ref{eq:gbath}), we get
\begin{equation}
\label{eq:ResIdentity}
\begin{split}
g^{(\pm)}(t)=\mathrm{Tr}_B\Big\{&1^{(+)} e^{-i \hatH_B^{(+)}t}1^{(+)}1^{(\pm)}\hatsigma^{(\pm)}(\beta)1^{(\pm)} \\
&\times 1^{(-)} e^{i \hatH_B^{(-)}t}1^{(-)}\Big\}
\end{split}
\end{equation}
from which standard algebra leads to
\begin{equation}
\begin{split}
\label{eq:gpm}
g^{(\pm)}(t)&=\mathrm{Tr}_B\{ e^{\mp 2i\hat{A}(\pm t)}\hatsigma(\beta) e^{\pm 2i\hat{A}(0)}\}\\
&\equiv\left\langle e^{\pm 2i\hat{A}(0)}e^{\mp 2i\hat{A}(\pm
t)}\right\rangle_{0}.
\end{split}
\end{equation}
Here the thermal density matrix of the bath is
\begin{equation}
\hatsigma(\beta)\equiv e^{-\beta \hatH_B}/\mathrm{Tr}_B\{e^{-\beta
\hatH_B}\},
\end{equation}
and $\hat{A}(t)=e^{i \hatH_Bt}\hat{A}e^{-i \hatH_Bt}$.
Since $\hatH_B$ corresponds to free bosons we can write the bath correlation function of Eq.\ (\ref{eq:gpm}) as
\begin{equation}
g^{(\pm)}(t)=e^{-W(\mp t)}
 \label{eq:linkedcluster}
\end{equation}
with the bosonic correlation function
\begin{equation}
W(t)\equiv  4\left[\langle \hat{A}^2(0)\rangle_0-\langle
\hat{A}(t)\hat{A}(0)\rangle_0\right].
\end{equation}
The correlation function can be evaluated using standard methods
from the field of quantum dissipative
systems\cite{Makhlin2001,Weiss2001} or using a Green's functions
approach.\cite{Braig2003,Flindt2004} Here, we just quote the
final result
\begin{equation}
\label{eq:W}
W(t)=\int_0^{\infty}d\omega
\frac{J(\omega)}{\omega^2}\{[1-\cos(\omega
t)]\coth\left[\frac{\beta\omega}{2}\right]+i\sin(\omega t)\},
\end{equation}
with
\begin{equation}
J(\omega)\equiv\sum_j|g_j|^2\delta(\omega-\omega_j)
\end{equation}
being the spectral function of the heat bath. In this work
we consider for simplicity Ohmic dissipation characterized by a coupling strength
$\alpha$ and a frequency cut-off $\omega_c$, such that the spectral function
reads
\begin{equation}
\label{eq:JOhm}
J_\Omega(\omega)=2\alpha \omega e^{-\omega/\omega_c}.
\end{equation}
For Ohmic dissipation, the correlation function is well known and reads\cite{Weiss2001}
\begin{equation}
\label{eq:Wexact}
W(t)=-2\alpha \ln\left[\frac{\left|\Gamma_E(1+\eta+it/\beta)\right|^2}{(1+i\omega_c t)\Gamma_E^2(1+\eta)}\right]\ ,
\end{equation}
where $\Gamma_E(x)$ is the Euler Gamma function and $\eta=1/\beta\omega_c$.

We consider energy scales and temperatures lower than the cut-off
$\omega_c$, such that $\eta\ll1$. In that limit, Eq.\
(\ref{eq:Wexact}) can be approximated as\cite{Martin2005}
\begin{equation}
\label{eq:Wapprox}
W(t)=2\alpha \ln\left[\frac{\sinh[i\pi\eta(1+i\omega_c t)]}{\sinh[i\pi\eta]}\right]\ ,
\end{equation}
using only elementary functions. We can then calculate
analytically the Laplace transform of the bath correlation
function. Using the integral identity
\begin{equation}
\begin{split}
\int_0^\infty dt e^{-zt}& \sinh(t+x)^{-y}=\\& \frac{2^y\  e^{-xy}}{z+y}
\ _2F_1\left[\frac{y+z}{2},y,1+\frac{y+z}{2},e^{-2x}\right]
\end{split}
\end{equation}
where $_2F_1(a,b,c,z)$ is the Gauss Hypergeometric function, we obtain
 \begin{equation}
\begin{split}
\label{eq:gzapprox}
g^{(\mp)}(z)=&\left(\frac{\beta}{\pi}\right) \frac{[1-e^{\mp i2\pi\eta}]^{2\alpha}}{2\alpha+\beta z/\pi}\\
&\times\
_2F_1\left[\alpha+\frac{\beta z}{2\pi},2\alpha,1+\alpha+\frac{\beta z}{2\pi},e^{\mp i2\pi\eta}\right]
\end{split}
\end{equation}
valid for $\alpha>0$ and $1/\beta\omega_c,|z|/\omega_c\ll1$.
Without coupling to the heat bath, we would have $W(t)=0$, and
from Eq.\ (\ref{eq:linkedcluster}) we would obtain
$g^{(\pm)}(z)=1/z$. The property of the Hypergeometric function,
$_2F_1(a,0,c,z)\equiv1$ for any value of $a,c$ and $z$, shows that
Eq.\ (\ref{eq:gzapprox}) indeed simplifies to this result in the limit
$\alpha\rightarrow 0$.

By Laplace transforming the bath-assisted hopping rates in Eq.\ (\ref{eq:bathassistedrates}), we finally find for real $z$
\begin{equation}
\begin{split}
\label{eq:Gammaz}
\Gamma_B^{(\pm)}(z)=&2T^2_c\int_0^{\infty}dt
e^{-zt}{\rm Re}\left[e^{-(i\varepsilon+\Gamma_R/2)t} g^{(\pm)}(t)\right]\\
=&T_c^2[g^{(\pm)}(z_+)+g^{(\mp)}(z_-)]
\end{split}
\end{equation}
with $z_\pm=z\pm i\varepsilon+\Gamma_R/2$. Here, we have used
the relation
\begin{equation}
[g^{(\pm)}(z)]^*=g^{(\mp)}(z^*), \label{eq:relation}
\end{equation}
for any complex $z$, which follows directly from the symmetry property
\begin{equation}
[W(t)]^*=W(-t)
\end{equation}
of the bosonic correlation function in Eq.\ (\ref{eq:W}) in combination with the expression in Eq.\ (\ref{eq:linkedcluster}).

Before concluding this appendix, we consider the regime, where the coupling to the heat bath is weak and the bath
temperature is high. Below, we show that the dynamics of the
DQD in that regime can be described by a charge detector model
with an effective dephasing rate
$\Gamma_d$.\cite{Leggett1987,Weiss2001,Makhlin2001} We derive the
dephasing rate starting from Eq.\ (\ref{eq:sigmaLR}). For weak
couplings, the bath remains unaffected by the electronic state of
the DQD, and we can perform a decoupling reading\cite{Braggio2009}
\begin{equation}
\label{eq:indepdecoupling}
\sigma_{ii}(t)\approx\rho_i(t)\otimes\sigma_\beta.
\end{equation}
Using this decoupling in
Eq.\ (\ref{eq:sigmaLR}) and tracing out the bath degrees of freedom we obtain
\begin{equation}
\label{eq:sigmaRLt}
\begin{split}
\mathrm{Tr}_B&\left\{\hatsigma_{LR}(t)\right\}=\\
&iT_c \int_0^t dt' e^{-(i\varepsilon+\Gamma_R/2)(t-t')}
g(t-t')
\left[\rho_{L}(t')-\rho_{R}(t')\right]
\end{split}
\end{equation}
having omitted the inhomogeneity entering Eq.\ (\ref{eq:sigmaLR}),
since we are only interested in long-time properties. The (single)
bath correlation function $g(t)$ is now
\begin{equation}
g(t)=\mathrm{Tr}_B\left\{e^{-i\hatH_B^{(+)}(t)}\hatsigma_\beta
e^{i\hatH_B^{(-)}(t)}\right\}.
\end{equation}
Using the polaron transformation in
Eq.\ (\ref{eq:PolaronS}), we readily find
\begin{equation}
\label{eq:gt}
g(t)=\left\langle e^{i\hat{A}(0)}e^{-2i\hat{A}(t)} e^{i\hat{A}(0)}\right
\rangle_{0}=e^{-\mathrm{Re}[W(t)]}
\end{equation}
where $W(t)$ is given in Eq.\ (\ref{eq:W}).

For the Ohmic bath, described by Eq.\ (\ref{eq:JOhm}), it is easy to demonstrate\cite{Makhlin2001}
in the long-time limit $\beta t\gg\hbar/2$ that $\mathrm{Re}[W(t)]\approx \Gamma_d  t$,
having defined the rate
\begin{equation}
\Gamma_d=2\alpha\pi k_B T.
\end{equation}
Following similar steps as those leading to Eq.\ (\ref{eq:bathassistedrates}) we find a bath-assisted hopping rate reading
\begin{equation}
\Gamma_B(t)\equiv2T_c^2\mathrm{Re}\left[e^{-(i\varepsilon+\Gamma_R/2+\Gamma_d)t}\right],
\label{eq:detectorassistedrates}
\end{equation}
or in Laplace space
\begin{equation}
\Gamma_B(z)=2T_c^2\frac{z+\Gamma_R/2+\Gamma_d}{\varepsilon^2+(z+\Gamma_R/2+\Gamma_d)^2}.
\label{eq:detectorassistedratesLP}
\end{equation}
Without coupling to the heat bath, we have $\Gamma_d=0$, and
the only broadening mechanism is the escape rate $\Gamma_R$ of electrons
to the right lead, which gives the hopping rate a width of
$\Gamma_R/2$. With weak coupling to the heat bath the rate is
additionally broadened by $\Gamma_d$, and the total dephasing rate
is $\Gamma_R/2+\Gamma_d$. This is similar to the results obtained
from a charge detector model with dephasing rate
$\Gamma_d$.\cite{Kiesslich2006,Gurvitz1997,Braggio2009}

\end{document}